\documentclass{aa} 
\usepackage[varg]{txfonts}

\usepackage{stmaryrd}

\begin{document}

\title{A Family of phase masks for broadband coronagraphy\\
  example of the wrapped vortex phase mask\\
  theory and laboratory demonstration}
\author{R. Galicher\inst{1}, E. Huby\inst{1}, P. Baudoz\inst{1},
  O. Dupuis\inst{1}}
\institute{LESIA, Observatoire de Paris, PSL Research University,
  CNRS, Sorbonne Universités, Univ. Paris Diderot, UPMC Univ. Paris
  06, Sorbonne Paris Cité, 5 place Jules Janssen, 92190 Meudon,
  France\\
\email{raphael.galicher at obspm.fr}
}
\authorrunning{R. Galicher et al.}
\titlerunning{Broadband coronagraphy with a wrapped vortex}
\date\today

 
  \abstract
{Future instruments need efficient
  coronagraphs over large spectral ranges to enable broadband imaging
  or spectral characterization of exoplanets that are $10^8$ times
  fainter than their star. Several solutions have been
  proposed. Pupil apodizers can attenuate the star intensity by a factor of 
  $10^{10}$ but they  only
  transmit a few percent of the light of the planet. Cascades of
  phase and/or amplitude masks can both attenuate
  the starlight and transmit most of the planet light, but the number
  of optics that require alignment makes this  solution impractical for an
  instrument. Finally, vector phase masks can be used to detect
  faint sources close to bright stars but they require the use of
  high-quality circular polarizers and, as in the previous solution,
  this leads to a complex instrument with numerous optics that require alignment and
  stabilization.}{We propose simple
  coronagraphs that  only need one scalar phase mask and one binary Lyot
  stop providing high transmission for the planet light
  ($>50\,\%$) and high attenuation of the starlight over a large
  spectral bandpass ($\sim30\,\%$) and a~$360^{\circ}$
  field-of-view.}{From mathematical considerations, we 
  find a family of 2D phase masks optimized for an unobscured
  pupil. One mask is an azimuthal wrapped vortex phase ramp. We probe its
  coronagraphic performance using numerical simulations and laboratory
  tests.}{From
  numerical simulations, we predict the wrapped vortex can attenuate the
  peak of the star image by a factor of~$10^4$ over a~$29\,\%$ bandpass
  and $10^5$ over a~$18\,\%$ bandpass with transmission 
  of more than $50\,\%$ of the planet flux at $\sim4\,\lambda/D$. We
  confirm these predictions in the laboratory in visible light
  between~$550\,$ and~$870\,$nm. We also obtain laboratory dark hole
  images in which exoplanets with fluxes that are $3.10^{-8}$ times
  the host star flux could be detected at~$3\,\sigma$.}
  {Taking advantage of a new technology for etching
  continuous 2D functions, a new type of mask can be easily
  manufactured opening up new possibilities for broadband coronagraphy.}


   \maketitle
%

   \section{Introduction}
Exoplanets and disks are objects that we study to
understand the formation of planetary systems. To date, more
than~$4  000$ exoplanets have been detected. Statistics of their
orbital parameters \citep{santos08,santos18} put constraints on the
models that explain the formation of planets
\citep{raymond14a,Izidoro18,santos17,dangelo18,Adibekyan19}. To put
more constraints on these models, we require detections of exoplanets that
orbit in the outer part of their system, close to the 
ice-line. However, the most commonly used detection techniques -- radial
velocity and transit -- do not easily observe such
exoplanets. Another challenge is the study of the exoplanet atmospheres
and surface temperatures, both of which require the measurement of the
planet spectrum in transmission, reflection, or emission. The transit
technique enables detections of molecules in the atmosphere of
exoplanets in favorable cases
\citep{snellen10,alonso19,essen19,espinoza19}, but transits
focus on exoplanets orbiting very close to their star. For nontransiting planets like $\beta-$Pictoris~b or HR\,8799\,e, medium- to
  high-resolution spectra have been used to detect molecules
  \citep{snellen14,lacour19} but such techniques can
  only be used to study known exoplanets. The discovery of planets in the outer
  part of their system therefore requires direct imaging techniques working
  on broad spectral bands. Statistical analyses of these detections put
  constraints on the models of planet formation. Imaging also allows
  spectral characterization that can be used to study exoplanet atmospheres.

Direct imaging is very challenging because exoplanets are up
to~$10^{10}$ times fainter than their host star in visible and
infrared. It requires dedicated instrumentation composed of extreme
adaptive optics systems and coronagraphs, as in the current
instruments Spectro-Polarimetric High-contrast Exoplanet REsearch
instrument at the Very Large Telescope \citep{beuzit19}, Gemini Planet Imager
\citep{macintosh14}, SCExAO/Subaru \citep{lozi18}, and Magellan adaptive
optics instrument \citep{close18}. These instruments have enabled the
detection of a few 
young giant exoplanets and numerous circumstellar disks of dust. However, none
of them can attenuate the star intensity by more than a
factor of~$10^3$ over a large spectral range for two
main reasons. Optical aberrations induce speckles in the science
image. Some of them are calibrated a posteriori using differential
imaging \citep{racine99,baba03,marois06} and in the near future their
intensity will be actively minimized thanks to focal plane wavefront
sensing \citep{jovanovic18}. The second limitation comes from the
current coronagraphs themselves because there is a trade-off between starlight attenuation and the spectral bandpass of the coronagraph.

For future instruments
that aim to characterize exoplanets such as Earth or mature
Jupiter, new broadband coronagraphs are required with high
attenuation of the starlight. Several solutions
have been proposed and a few of them were validated in the laboratory,
attenuating the starlight by a factor of~$10^8$ to~$10^9$
\citep{galicher11,mawet11,serabyn19}. No solution is perfect and
trade-offs remain. For example, for the shaped pupil
coronagraphs, a comprimise must be made between the field of view, the
attenuation of the starlight, and the transmission of the exoplanet
flux \citep{cady17}.
In this paper, we present a family of coronagraphs based on  only one scalar
phase mask. These are efficient over a $30\,\%$ spectral band
with a very high transmission of the planet intensity in a~$360^{\circ}$
field of view. In section~\ref{sec:theory}, we briefly review the
current limitations of phase mask coronagraphy and we use a
mathematical approach to derive the phase functions of a family of
scalar broadband coronagraphic masks. One of these masks is the
wrapped vortex phase mask that we study using numerical simulations in
section~\ref{sec:simu}. We then present our wrapped vortex prototype
in section~\ref{sec:manufacturing} and its laboratory performance in
section~\ref{sec:labo}.

\section{Theory of wrapped broadband phase masks}
\label{sec:theory}

\subsection{Phase mask for coronagraphy}
\begin{figure}[!ht]
  \centering
  \includegraphics[width=6cm]{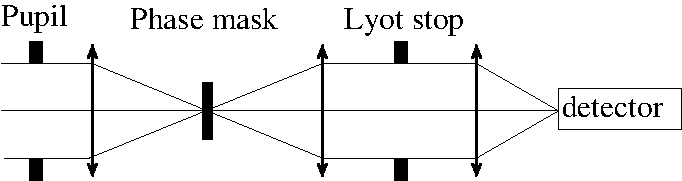}
  \caption{Principle of a coronagraph. The on-axis source image is
    centered on a focal plane phase mask that scatters the light
    outside the geometrical pupil in the following plane. The Lyot
    stop filters out the starlight that does not reach the detector.}
  \label{fig:coro_princ}
\end{figure}

We consider a coronagraph composed of a single focal plane phase mask
followed by a classical binary Lyot stop
(figure~\ref{fig:coro_princ}).
The star image is centered on the mask. We call~$r$ and~$\theta$ the
polar coordinates in this plane. If the phase
function~$G(r,\,\theta)$ is well chosen, all the starlight is 
rejected outside of the geometrical pupil in the following pupil
plane and it is stopped by the binary Lyot stop. The starlight is
filtered out from the system and  does not reach the
detector. The light of an off-axis source is not focused on the center
of the phase mask; it goes through the system, reaches the
detector, and its image is detected. Designing such a coronagraph requires finding the 2D phase function~$G(r,\,\theta)$ that maximizes both
the starlight attenuation and the transmission of an off-axis
source. Numerous phase functions that theoretically stop~$100\,\%$ of
the starlight have been proposed
\citep[etc.]{Rouan00,Mawet05,murakami08}.

However, it is  difficult to fabricate a component that adds the 
  perfect phase~$G_0$ at all wavelengths regardless of the type of mask:
transmissive or reflective scalar masks or vector phase mask.
A transmissive scalar phase mask is made of a material whose
thickness~$e(r,\,\theta)$ varies in the field of view. The variations
in thickness induce variations in the phase
shift~$G(r,\,\theta)=2\,\pi\,(n(\lambda)-1)\,e(r,\,\theta)/\lambda$
with~$n(\lambda)$ being the optical index of the material at
wavelength~$\lambda$. The thickness~$e$ can therefore be chosen so that the
phase~$G$ is equal to the  `perfect'~$G_0$ at one
wavelength~$\lambda_0$ which we henceforth refer to as the optimized wavelength. At all other wavelengths, the phase function~$G$
differs from the  perfect function~$G_0$ and the starlight is not
totally filtered out by the coronagraph. Such a phase mask is
therefore monochromatic. A reflective phase mask is also based on
variations of the thickness~$e(r,\,\theta)$ inducing variations of the
phase~$G(r,\,\theta)=4\,\pi\,e(r,\,\theta)/\lambda$. As for the
transmissive mask, the function~$G$ is equal to the  perfect
function~$G_0$ at one wavelength~$\lambda_0$ only.

To mitigate the chromatic dependence of scalar masks,
\citet{galicher11} and \citet{mawet11} proposed the use of several
monochromatic phase masks in cascade. This concept enables  the starlight to
be attenuated over~$\sim20\,\%$ of the spectral bandpass but both
of these latter two groups of authors concluded that aligning two or
three coronagraphs in cascade is too demanding in terms of instrument
stability.

Other phase masks are vectorial masks \citep{mawet09}, which use a
half-wave plate whose 
neutral axes rotate in the field of view so that the mask induces a
geometrical phase shift~$G(r,\,\theta)$. However, the phase shift is
polarization dependent and is not perfect because of
retardance errors (chromaticity and errors on the orientation of
neutral axes). Therefore, the phase function~$G$ differs from the 
  perfect function~$G_0$ and part of the starlight leaks through the
coronagraph. It has been shown that the leakage can be minimized if
the vector phase mask is set between circular
polarizers~\citep{serabyn19}. However, this  removes half of the
putative exoplanet flux (removing one of the two polarization states)
and requires the use and alignment of polarizers, which is
not the optimal solution in terms of instrument stability.

\subsection{General formalism for scalar azimuthal phase masks}
\label{subsec:gene_form}
To reduce the complexity of the instrument and maintain the high performance of the
coronagraph, we consider a scalar phase mask. Furthermore, to mitigate the wavelength dependance of the phase shift induced by a
scalar mask, we look for a~$G(r,\,\theta,\,\lambda)$ function that is
optimized to cancel the starlight over a broadband. In the rest of the
paper, we assume that the optical index~$n$ of the mask material does
not vary with wavelength. For example, the optical index of
fused silica is constant at~$0.3\,\%$ between~$600\,$
and~$800\,$nm. We also assume an azimuthal phase mask, meaning the
phase function~$G$ does not vary with the radial coordinate~$r$:
\begin{equation}
G(\theta,\,\lambda)=\frac{2\,\pi\,\alpha\,e(r,\,\theta)}{\lambda}
= G_0(\theta)\,\frac{\lambda_0}{\lambda}
.\end{equation}
The parameter~$\alpha$ is a constant that is equal to~$n-1$ for a
transmissive mask and~$2$ for a reflective mask. The function~$G_0$
is the `perfect' phase function that is induced at the optimized
wavelength~$\lambda_0$.

Considering a full entrance pupil (no spider or obscuration),
\citet{ma12} expressed the amplitude of the stellar electric
field in the Lyot plane as a sum of Hankel transforms of Bessel
functions. These latter authors also demonstrated that if the~$G_0$ function is
$\pi$-periodic, the sum reduces to a single term proportional
to:
\begin{equation}
c_0(\lambda) = \frac{1}{2\,\pi}\int_0^\pi\exp{\left(i\,G_0(\theta)\,
  \lambda_0/\lambda\right)}\,\mathrm{d}\theta
  .\end{equation}

In the rest of the paper, we focus on~$\pi$-periodic~$G_0$ functions
that minimize~$c_0(\lambda)$ -- i.e. the starlight leakage -- over a large
spectral bandpass centered on the optimized wavelength $\lambda_0$. To
ensure a weak chromatic dependence of the coronagraphic performance,
\citet{ma12} looked for~$G_0$ functions that cancel out~$c_0$ and its
  first derivative at~$\lambda_0$. We go one step further and force the
second derivative to be null too:
\begin{equation}
  \left\{
  \begin{aligned}
    c_0(\lambda_0) &=&0\\
  \frac{\partial c_0}{\partial\lambda}(\lambda_0)&=&0\\
  \frac{\partial^2c_0}{\partial\lambda^2}(\lambda_0)&=&0.\\
  \end{aligned}
  \right.
\label{eq:systder}
\end{equation}
Using the following expressions for the derivatives,
\begin{equation}
  \left\{
  \begin{aligned}
  \frac{\partial c_0}{\partial\lambda}(\lambda)&=
  -\frac{i\,\lambda_0}{2\,\pi\,\lambda^2}\int_0^\pi
  G_0(\theta)\,e^{i\,\lambda_0\,G_0(\theta)/\lambda}\,\mathrm{d}\theta\\
    \frac{\partial^2 c_0}{\partial\lambda^2}(\lambda)&=
  \frac{i\,\lambda_0}{\pi\,\lambda^3}\int_0^\pi
  \left[G_0(\theta) + \frac{i\lambda_0\,G_0^2(\theta)}{2\,\lambda}
    \right]\,e^{i\,\lambda_0\,G_0(\theta)/\lambda}\,\mathrm{d}\theta,\\
  \end{aligned}
  \right.
\end{equation}
one can rewrite~Eq.~\ref{eq:systder}:
  \begin{subequations}
  \begin{align}
    \int_0^\pi \exp{\left(i\,G_0(\theta)\right)}\,\mathrm{d}\theta
    =0\label{eq:c0}\\ 
    \int_0^\pi
    G_0(\theta)\,\exp{\left(i\,G_0(\theta)\right)}\,\mathrm{d}\theta
    =0\label{eq:dc0}\\ 
    \int_0^\pi
    G_0^2(\theta)\,\exp{\left(i\,G_0(\theta)\right)}\,\mathrm{d}\theta
    =0.\label{eq:d2c0}
\end{align}
\end{subequations}

  One family of $G_0$ functions that obey Eq.\,\ref{eq:c0} is defined
  by
  \begin{equation}
    \forall \theta\in\left[\frac{\pi}{2},\pi\right[,
  \exists\,\theta_1\in\left[0,\frac{\pi}{2}\right[, G_0(\theta) =
  G_0(\theta_1)+(2\,k_\theta+1)\,\pi,
  \label{eq:cond1}
  \end{equation}
  with~$k_\theta$ in~$\mathbb{Z}$. The four quadrant phase mask
  \citep[FQPM,][]{Rouan00} 
  obeys this equation with $k_\theta=0$ and
  $G_0(\theta)=0$ for~$\theta\in\left[0,\pi/2\right[$. The eight octant 
   \citep[EOPM,][]{murakami08} and the six level phase masks
  \citep[SLPM,][]{hou14} are also part of the family if we
  allow~$k_\theta$ to vary with $\theta$ ($k_\theta=-1$ or
  $k_\theta=0$). 
  
  We consider the family of functions defined by
  condition~\ref{eq:cond1} with $k_\theta=k$ not varying with
  $\theta$. We then search for~$G_0$ functions that also obey
  Eq.\,\ref{eq:dc0} that can be written
  \begin{equation*}
  \int_0^{\pi/2}
  G_0(\theta)\,\exp{\left(i\,G_0(\theta)\right)}\,\mathrm{d}\theta +
  \int_{\pi/2}^{\pi}
  G_0(\theta)\,\exp{\left(i\,G_0(\theta)\right)}\,\mathrm{d}\theta =0.
  \end{equation*}
  We account for condition~\ref{eq:cond1} and simplify the result to:
  \begin{equation*}
    (2\,k+1)\,\pi\,\int_0^{\pi/2}\exp{\left(i\,G_0(\theta)\right)}
    \,\mathrm{d}\theta =0.
  \end{equation*}
  This equation is validated by the family of functions such
  as
  \begin{equation}
  \forall \theta\in\left[\frac{\pi}{4},\frac{\pi}{2}\right[,
  \exists\,\theta_1\in\left[0,\frac{\pi}{4}\right[, G_0(\theta) =
  G_0(\theta_1)+(2p_\theta+1)\,\pi,
  \label{eq:cond2}
  \end{equation}
  with~$p_\theta$ in~$\mathbb{Z}$. Unlike the FQPM and the SLPM, the
  EOPM obeys this condition with~$p_\theta=0$ and~$G_0(\theta)=0$
  for~$\theta\in\left[0,\pi/2\right[$.
  
  Finally, we search for a sub-family that also obeys
  Eq.\,\ref{eq:d2c0} imposing a uniform $p_\theta=p$. One can
  transform Eq.\,\ref{eq:d2c0} into
  \begin{eqnarray*}
    && \int_0^{\pi/4}
    G_0^2(\theta)\,\exp{\left(i\,G_0(\theta)\right)}\,\mathrm{d}\theta\\
    -&&\int_{0}^{\pi/4}
    \left[G_0(\theta)+(2\,p+1)\,\pi\right]^2\,\exp{\left(i\,G_0(\theta)\right)}\,\mathrm{d}\theta\\
    -&&\int_{0}^{\pi/4}
    \left[G_0(\theta)+(2\,k+1)\,\pi\right]^2\,\exp{\left(i\,G_0(\theta)\right)}\,\mathrm{d}\theta\\
    +&&\int_{0}^{\pi/4}
    \left[G_0(\theta)+(2\,k+2\,p+2)\,\pi\right]^2\,\exp{\left(i\,G_0(\theta)\right)}\,\mathrm{d}\theta\\
    &&=0.\\
  \end{eqnarray*}  
  Using Eqs.\,\ref{eq:c0} and \ref{eq:dc0} and
  conditions~\ref{eq:cond1} and~\ref{eq:cond2}, one finds
  \begin{equation*}
    \left(4\,k\,p+2\,p+2\,k+1\right)\int_0^{\pi/4}
  \,\exp{\left(i\,G_0(\theta)\right)}\,\mathrm{d}\theta=0.
\end{equation*}
This equation is validated if and only if the integral is null and one
family of solutions is
\begin{equation}
  \forall \theta\in\left[\frac{\pi}{8},\frac{\pi}{4}\right[,
  \exists\,\theta_1\in\left[0,\frac{\pi}{8}\right[, G_0(\theta) =
  G_0(\theta_1)+(2\,m_\theta+1)\,\pi,
  \label{eq:cond3}
  \end{equation}
with~$m_\theta$ in~$\mathbb{Z}$. This parameter can depend on~$\theta$
although we
do not use this property in the rest of the paper (we
write~$m_\theta=m$). We note that the EOPM does not validate
Eq.\,\ref{eq:cond3}.

  Finally, $G_0$ functions that obey the three
  conditions\,\ref{eq:cond1}, \ref{eq:cond2}, and \ref{eq:cond3} also 
  obey condition \ref{eq:systder} meaning that the coronagraphic performance is
  perfect at the optimized wavelength~$\lambda_0$ ($c_0(\lambda_0)=0$)
  and is not very sensitive to chromatism (the two first derivatives
  of~$c_0$ are null at~$\lambda_0$). For example, given~ $f,$ a
  function defined over $[0,\pi/8[$ and $m$, $p$, and $k,$ three
      numbers in $\mathbb{Z}$, one family of solutions is
  \begin{equation}
  G_0(\theta)=
    \begin{cases}
       f(\theta)&0\le\theta<\frac{\pi}{8}\\
       f(\theta-\frac{\pi}{8})+\left[2m+1\right]\pi&\frac{\pi}{8}\le\theta<\frac{\pi}{4}\\
       f(\theta-\frac{\pi}{4})+\left[2p+2m+2\right]\pi&\frac{\pi}{4}\le\theta<\frac{3\pi}{8}\\
       f(\theta-\frac{3\pi}{8})+\left[2p+1\right]\pi&\frac{3\pi}{8}\le\theta<\frac{\pi}{2}\\
       f(\theta-\frac{\pi}{2})+\left[2k+2m+2\right]\pi&\frac{\pi}{2}\le\theta<\frac{5\pi}{8}\\
       f(\theta-\frac{5\pi}{8})+\left[2k+1\right]\pi&\frac{5\pi}{8}\le\theta<\frac{3\pi}{4}\\
       f(\theta-\frac{3\pi}{4})+\left[2k+2p+2\right]\pi&\frac{3\pi}{4}\le\theta<\frac{7\pi}{8}\\
       f(\theta-\frac{7\pi}{8})+\left[2k+2p+2m+3\right]\pi&\frac{7\pi}{8}\le\theta<\pi\\
      G_0(\theta-\pi) &\pi\le\theta<2\pi.
    \end{cases}
    \label{eq:defG}
\end{equation}
 We note that we can reorder the values $G_0(\theta)$ inside each
 interval $[n\,\pi/8,(n+1)\,\pi/8[$ with $n$ in
 $\llbracket0,7\rrbracket$ as long as conditions \ref{eq:cond1},
 \ref{eq:cond2}, and~\ref{eq:cond3} are validated.

 \subsection{Wrapped vortex phase mask}
 \label{subsec:theory_wrapped}
\begin{figure}[!ht]
  \centering
  \includegraphics{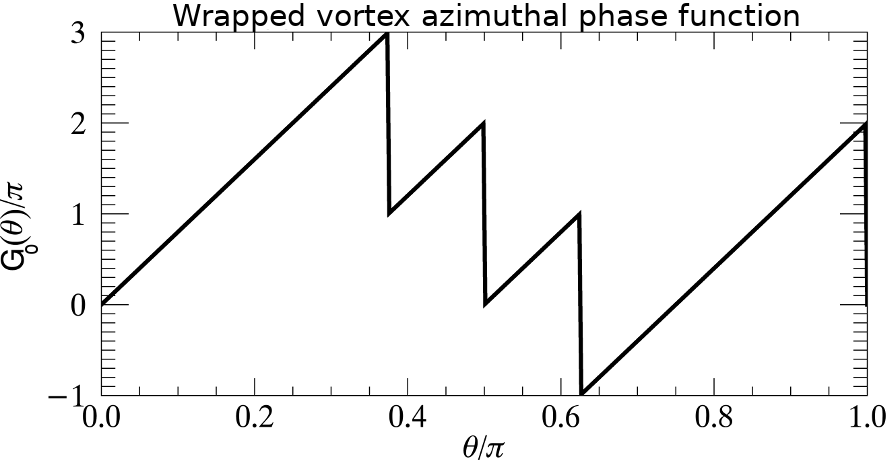}
  \caption{Wrapped vortex ramp~$G_0$ defined by Eq.\,\ref{eq:defGex}.}
  \label{fig:1D}
\end{figure}

\begin{figure}[!ht]
  \centering
  \includegraphics[width=5cm]{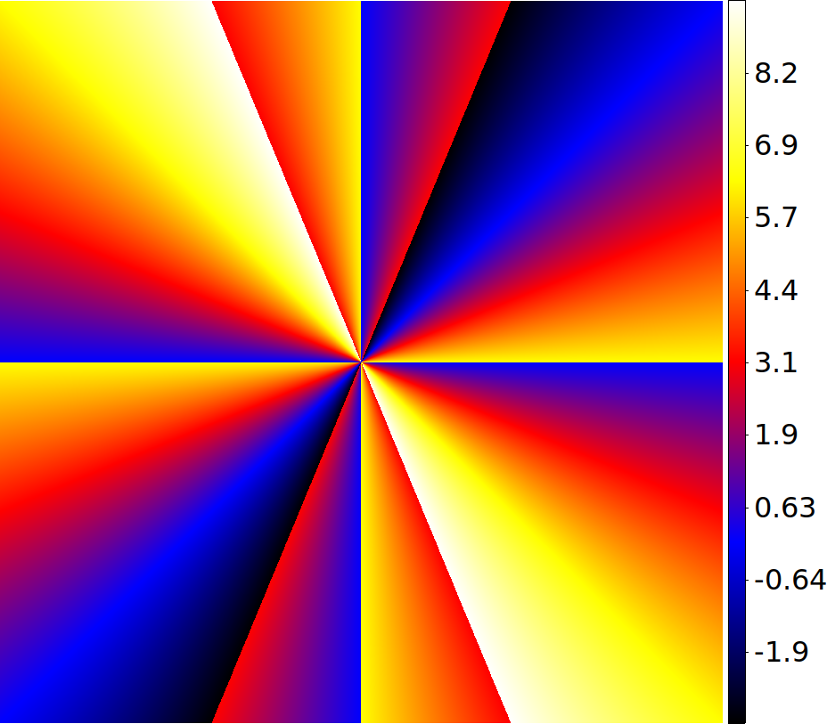}
  \caption{Two-dimensional map of the wrapped vortex defined by
    Eq.\,\ref{eq:defGex}. The color bar gives the phase in radians.}
  \label{fig:2D}
\end{figure}

 Equation\,\ref{eq:defG} defines an infinite number of
 functions. Here, we use an additional condition:
 avoiding~$\pi$ discontinuities. Such $\pi$ transitions that exist in
 the FQPM, SLPM, and EOPM strongly affect the image of off-axis sources
 reducing the effective field of view. One example of $G_0$ that obeys
 Eq.\,\ref{eq:defG} with no $\pi$ discontinuities is defined by $m=0$,
 $p=0$, $k=-1$ and $f(\theta)=8\,\theta$:
 \begin{equation}
  G_0(\theta)=
    \begin{cases}
      8\,\theta&\ 0\le\theta<\frac{3\pi}{8}\\
       8\,\theta-2\,\pi&\ \frac{3\pi}{8}\le\theta<\frac{\pi}{2}\\
       8\,\theta-4\,\pi&\ \frac{\pi}{2}\le\theta<\frac{5\pi}{8}\\
       8\,\theta-6\,\pi&\ \frac{5\pi}{8}\le\theta<\pi\\
      G_0(\theta-\pi) &\ \pi\le\theta<2\pi.
    \end{cases}
    \label{eq:defGex}
 \end{equation}
This function is continuous everywhere except at positions $\theta=0$,
$3\pi/8$, $\pi/2$, $5\pi/8$, $\pi$, $11\pi/8$, $3\pi/2$, $13\pi/8$
where there are $2\,\pi$ steps (figures\,\ref{fig:1D} and\,\ref{fig:2D}).
The resulting phase mask that we refer to as the wrapped vortex of topological
charge~$8$ is a phase ramp that would go from~$0$ to~$16\,\pi$ but
that is wrapped eight times.

Replacing~$f(\theta)=8\,\theta$ by~$f(\theta)=0$, Eq.\,\ref{eq:defGex}
describes one mask derived from Diophantine optics \citep[fig.\,3
  in][]{rouan16}. In future work, we will study how Diophantine
optics can be linked to the family of functions described by
Eq.\,\ref{eq:defG}.

\section{Numerical simulations of a wrapped vortex}
\label{sec:simu}
\subsection{Coronagraphic attenuation}
\label{subsec:att}
We run numerical simulations to estimate the coronagraphic attenuation
of the central star using the wrapped vortex defined
by~Eq.\,\ref{eq:defGex}. We assume Fourier optics and use fast
  Fourier transforms to propagate light from a pupil plane to a
  focal plane and vice versa. The entrance
pupil is a circular disk of~$256\,$pixels in arrays
of~$16\,384\,$pixels. The phase mask is set in the
first focal plane. The Lyot diaphragm is in the following pupil
plane. Eventually, the image is recorded in the last focal plane.

We assume the
phase mask is made of a single 
piece of material that is etched to induce the perfect phase
shift~$G_0$ at the optimized wavelength~$\lambda_0$. We
neglect the variation of the optical index with wavelength (see
section~\ref{subsec:gene_form}). We assume a Lyot stop whose
diameter~$D$ is~$95\,\%$ of the entrance pupil diameter.
\begin{figure}[!ht]
  \centering
  \includegraphics[width=7.5cm,height=6cm]{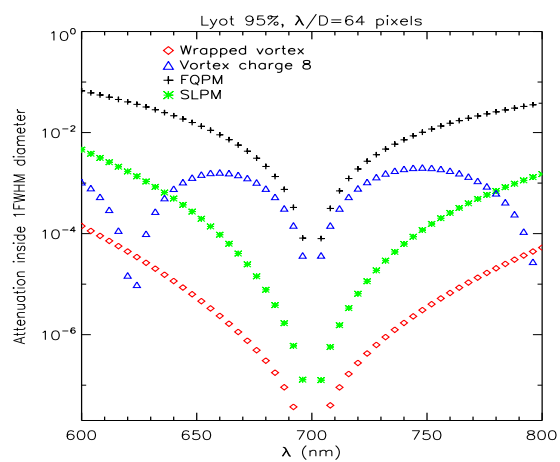}
  \caption{Numerical simulations: Coronagraphic attenuation
    integrated inside a disk of~$1\,\lambda/D$-diameter centered on
    the optical axis for a four quadrant phase mask
    (black plus), a vortex phase mask of topological charge $8$ (blue
    triangle), a six-level phase mask (green stars), and a
    wrapped-vortex phase mask defined by Eq.\,\ref{eq:defGex} (red
    diamond).}
  \label{fig:attenwrappedvortex}
\end{figure}
For each wavelength, we calculate the integrated energy inside
$1\,\lambda/D$-diameter at the center of the coronagraphic image. We
normalize this energy by the energy that is calculated in the same way
using the noncoronagraphic image (i.e., $G_0(\theta)=0$ everywhere) at
the same wavelength. The result is the attenuation~$A$ that is plotted
in~figure\,\ref{fig:attenwrappedvortex}. By definition, $A$ is equal to~$1$
at all wavelengths if no coronagraph is used. We compare the wrapped
vortex to FQPM, the vortex phase mask of topological charge~$8$,
and SLPM. All masks are optimized at~$\lambda_0 = 700$\,nm.

As expected from \citet{hou14} and as demonstrated in laboratory by
\citet{patru18}, SLPM is less sensitive to chromatism than~FQPM. The
attenuation for the vortex of charge~$8$ is minimum at
$\lambda_0$, $\lambda_1= 16/18\,\lambda_0 = 622$\,nm (equivalent to a
vortex of charge~$9$) and at $\lambda_2 =16/14\,\lambda_0 =
800$\,nm (equivalent to a vortex of charge~$7$). The attenuations
at~$\lambda_1$ and~$\lambda_2$ are not infinite because the charge is
odd. They are large however because the charge is
high~(\citet{Mawet05}). The wrapped vortex attenuates the starlight up
to approximately 10 times better than the~SLPM and up to approximately $500$ times
better than the~FQPM over the $\sim29\,\%$ bandpass ($600$ to
$800$\,nm). Except for short bandpasses around $\lambda_1$ and
$\lambda_2$, the wrapped vortex attenuation is also~$10$ to~$100$
times better than for the vortex of charge~$8$. Finally, on average,
the wrapped vortex attenuates the starlight by a factor of~$10^4$ over
a~$29\,\%$ bandpass, $10^5$ over a~$18\,\%$ bandpass, and~$10^6$ over
a~$10\,\%$ bandpass.

\subsection{Off-axis transmission}
\label{subsec:wrapped_offpsf}
In section~\ref{subsec:att}, we show that the wrapped vortex phase
mask can attenuate the central starlight over a large bandpass.
\begin{figure}
  \centering
  \includegraphics[width =.49\textwidth]{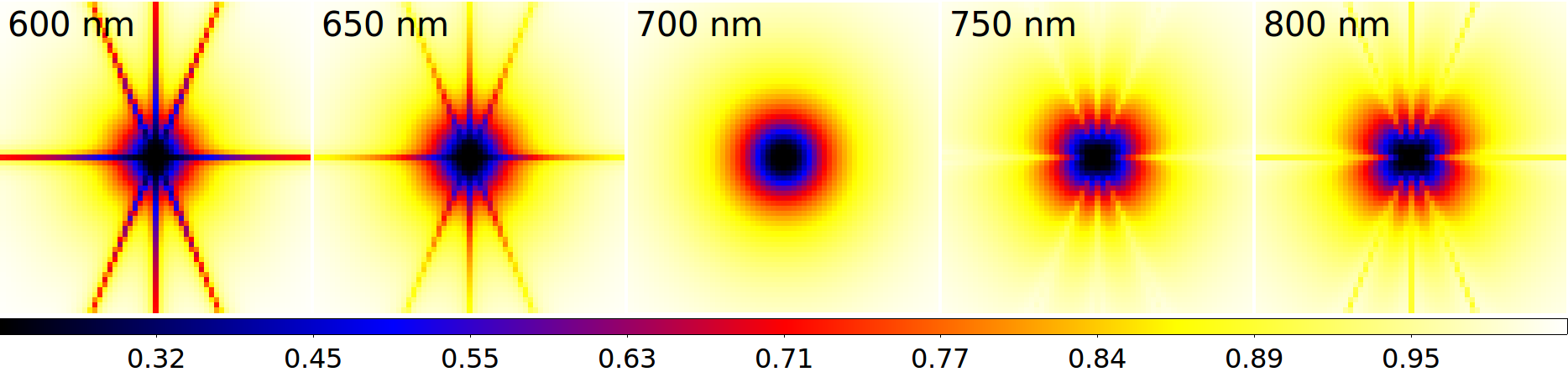}
  \caption{Numerical simulations: Maps of the energy transmission for
    a point-like source 
    observed through a coronagraph using a wrapped vortex phase mask
    at wavelengths~$\lambda$ from $600$\, to
    $800$\,nm. The excursion of the source goes from $-30\,\lambda/D$ to
    $30\,\lambda/D$ from the center of the phase mask in both
    horizontal and vertical directions. The color bar goes from $0$ (no
    light is transmitted) to~$1$ (all flux is transmitted).}
  \label{fig:transmission_map}
\end{figure}
The second role of a coronagraph is to maximize the transmission of
the planet light through the instrument. Therefore, we study the
transmission of the coronagraph for an off-axis source (e.g., the
planet). The transmission is the ratio between the maximum of the
image of off-axis sources obtained with the phase mask and without
the mask (i.e.,~$G_0(\theta)=0$). We calculate the transmission at
several wavelengths~$\lambda$ for sources angularly separated
from~$-30\,\lambda/D$ to~$30\,\lambda/D$ from the optical axis with
steps of~$1\,\lambda/D$. Maps are shown in
figure~\ref{fig:transmission_map} for five wavelengths from~$600\,$
to~$800\,$nm. At~$700\,$nm, the map is
centro-symmetrical because the steps of
material on the phase mask induce~$2\,\pi$ phase shifts. At this
wavelength, the inner working angle at which~$50\,\%$ of the planet
flux is transmitted is~$\sim4\,\lambda/D$. If the wavelength is not
equal to~$700$\,nm, the steps of materials 
induce phase shifts that are not~$2\,\pi$ and interference modifies the
transmitted flux. Therefore, the transmission map is not
centro-symmetrical. Such inhomogeneous
transmission over the field of view requires careful calibrations
to extract the photometry of a planet or of a circumstellar disk. However, we measure that over a~$100$\,nm bandpass around~$\lambda_0=700\,$\,nm
($14\,\%$ bandpass), the transmission varies by less than~$10\,\%$.

\begin{figure}[!ht]
  \centering
  \includegraphics[width=.45\textwidth]{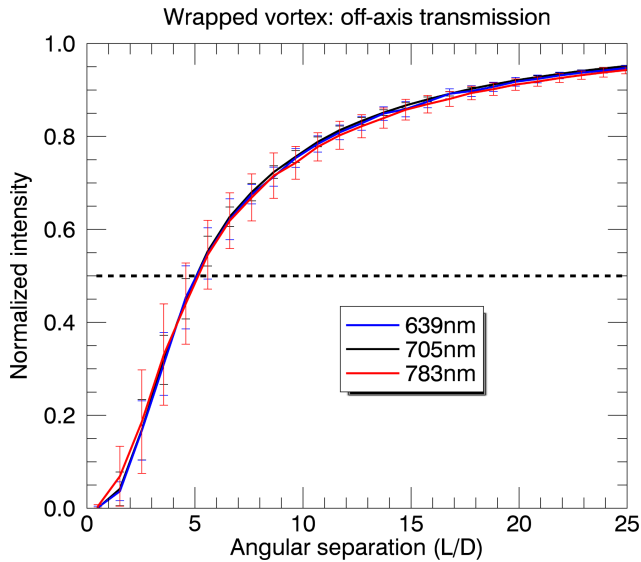}
  \caption{Numerical simulations: Azimuthal average of the energy
    transmission for a point-like source as a function of the angular
    separation from the optical axis for a wrapped vortex phase mask
    optimized at~$700\,$nm. The transmissions are calculated at
    $639\,$nm (blue), $705\,$nm (black), and $783\,$nm (red). The
    error bars show the standard deviation computed azimuthally.}
  \label{fig:transmission_curve}
\end{figure}

\begin{figure}[!ht]
  \centering
  \includegraphics[width=.45\textwidth]{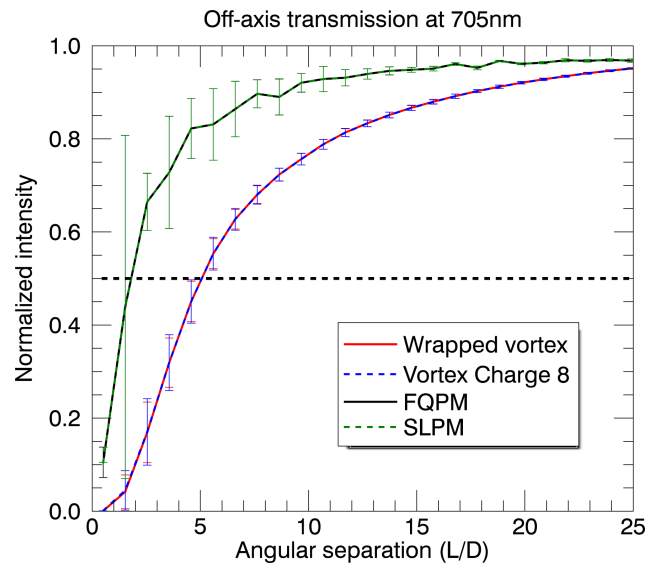}
  \caption{Numerical simulations: Same as
    figure~\ref{fig:transmission_curve} at~$705\,$nm but for several phase
    masks optimized at~$700\,$nm.}
  \label{fig:transmission_curve_allcoro}
\end{figure}

In figure~\ref{fig:transmission_curve}, we plot the 
azimuthal average of the energy transmission for a point-like source
at $639\,$, $705\,$, and $783\,$nm as a function of the angular separation
from the optical axis for a wrapped vortex phase mask optimized
at~$700\,$nm.  The three wavelengths are the ones of the laser diodes
that we use in the experiment (see section~\ref{subsec:exp_att}). The
    error bars show the standard deviation computed azimuthally.
We find that the transmission curves are very similar with an inner
  working angle of~$\sim5\,\lambda/D$.  For comparison, we plot in
  figure~\ref{fig:transmission_curve_allcoro} the same curve
  at~$705\,$nm for phase masks made of a single piece of material:
  FQPM, SLPM, vortex of charge~8, and wrapped vortex. The masks are
  optimized at~$700\,$nm. As already
  demonstrated in \citet{patru18}, SLPM and FQPM transmission curves
  are the same with an inner working angle of~$\sim2\,\lambda/D$. The
  large error bar at~$2\,\lambda/D$ comes from the strong variation of
  the transmission in the field of view because of
  the~$\pi$-transitions of these masks. We
  also find that the transmission curves are the same for the
  wrapped vortex and the vortex of charge~8 with a~$\sim5\,\lambda/D$
  inner working angle. However, unlike the wrapped vortex, we find
  from numerical simulations (not plotted here) that the transmission
  curve slightly evolves with wavelength for the scalar vortex of
  charge~8.

\section{Manufacturing of wrapped vortex mask}
\label{sec:manufacturing}
\subsection{Specifications}
\label{subsec:spec}
We ordered one wrapped vortex component from the Zeiss company  whose phase function is described
by~Eq.\,\ref{eq:defGex}. The component is a square
of~$17.5^{+0}_{-0.1}\,$mm$\,\times\,17.5^{+0}_{-0.1}\,$mm made of fused silica
with a thickness of~$1\,$mm. The parallelism between the two faces of
the substrate is smaller than~$20\,$arcseconds. The surface quality
before ion-etching is better than~$\lambda/10$ peak-to-valley
at~$633\,$nm in transmission for spatial frequencies larger
than~$0.05\,$cy/mm. The wrapped ramp of material thickness is
calculated so that the phase function obeys~Eq.\,\ref{eq:defGex} with
an accuracy of~$\pm5\,\%$ at the optimized wavelength of~$700\,$nm. It was
created by ion-etching -- the manufacturer can etch continuous
2D functions. After the ramp is etched, both faces are coated so that
the average reflection coefficient in amplitude is below~$0.2\,\%$ per
surface between $600\,$ and~$800\,$nm.

\subsection{Delivered component}
\label{subsec:delivered_component}
\begin{figure}[!ht]
  \centering
  \includegraphics{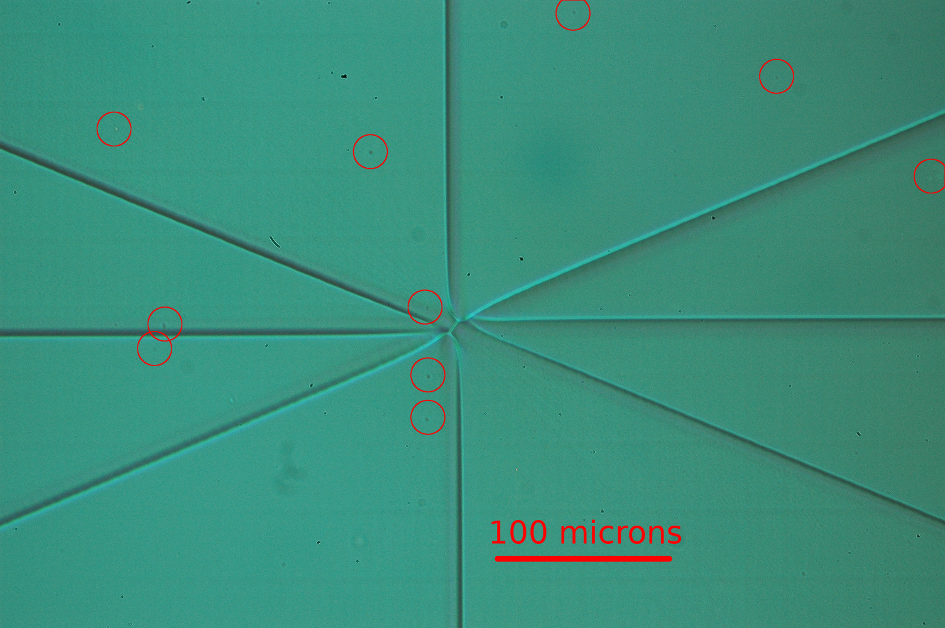}
  \caption{Image of the wrapped vortex component obtained with an
    optical microscope. Dusts and defects on the component are
    encircled in red. The other defects are on the microscope optics.}
  \label{fig:mask2pos7_optmicro}
\end{figure}

All specifications are respected by the manufacturer but the
optimized wavelength is larger than expected. This is because we
asked the manufacturer to fabricate different phase masks on the same
wafer. Therefore, it was complicated to constrain the optimal depth of
etching and the different 2D phase functions simultaneously. Confocal
measurements provided by the manufacturer find that the step of
material near the border of the mask (meaning at~$\sim8.7\,$mm from
the center) is about~$10\,\%$ higher than the specification. Assuming
the step of material has the same height at the center of the mask, we
can expect the mask to be optimized at~$770\,$nm instead of the
specified~$700\,$nm.
Figure\,\ref{fig:mask2pos7_optmicro} shows a visible microscope image
of the component.
Most of the detected spots are defects on the microscope optics. Only
a little dust and a few defects are present on the component (encircled in
red). In the following sections, the diameter of the point spread function
that is focused on the phase mask is between~60 and~$100\,\mu$m
depending on the wavelength. Therefore, only the three defects that
are the closest to the center of the mask are lit up. As they are
very small in size ($<0.4\,\mu$m), we expect their effect to be
negligible. From the image, we also measure the width of the
$2\,\pi$-phase steps to be between approximately~$5\,$and~$8\,\mu$m. This is quite
large with respect to what can be fabricated for binary masks
like~FQPM \citep[$<0.5\,\mu$m in][]{bonafous16}. However, unlike
the~FQPM, numerical simulations predict that the wrapped vortex gives
good coronagraphic attenuations even with transition widths
of~$\sim10\,\%$ of~$\lambda/D$. We believe that this is because the phase
steps are $2\,\pi$ in phase instead of~$\pi$ for the~FQPM.

\section{Laboratory performance of a wrapped vortex}
\label{sec:labo}
\subsection{THD2 bench}
\label{subsec:thd2}
\begin{figure*}[!ht]
  \centering
  \includegraphics[width=.85\textwidth]{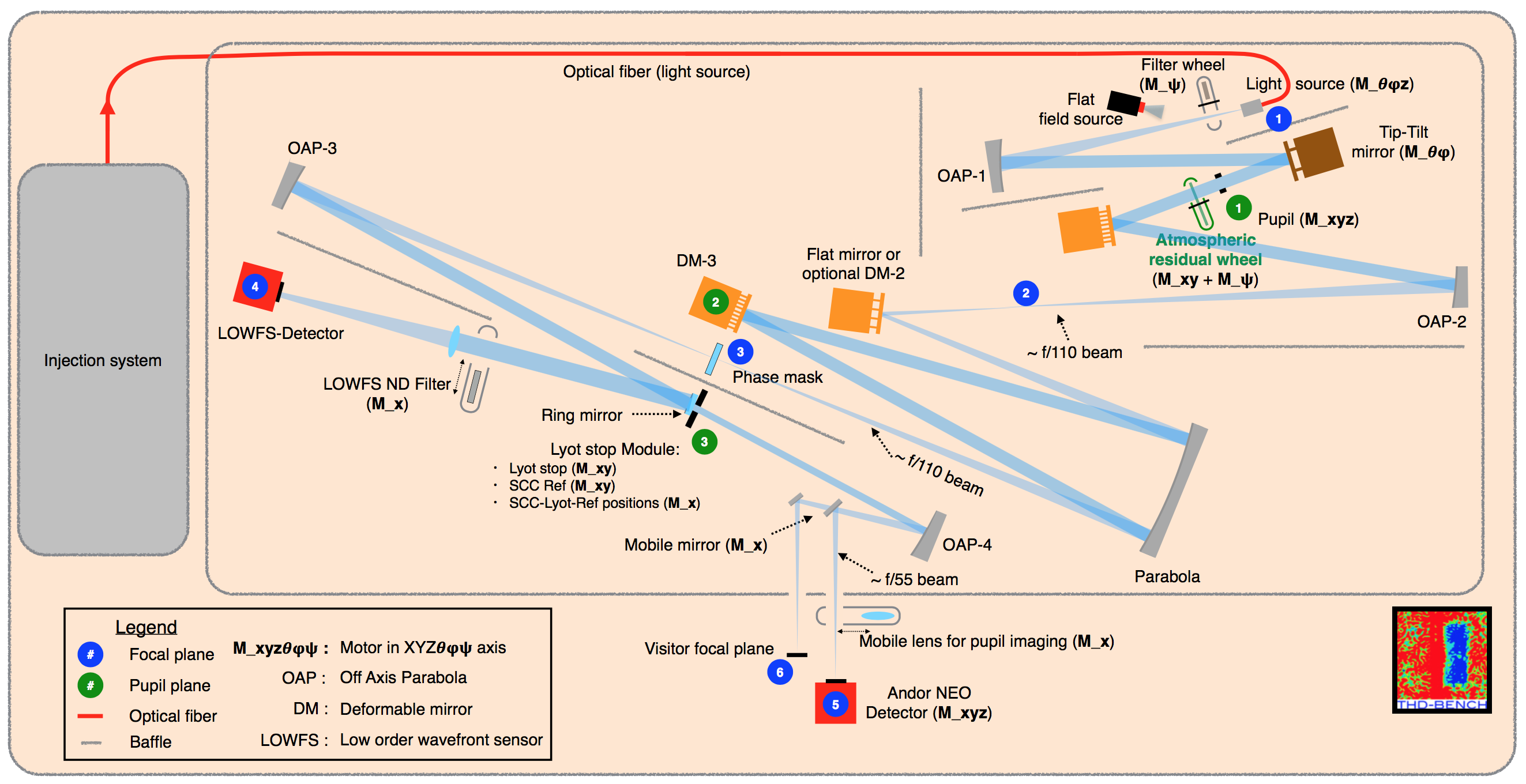}
  \caption{Layout of the THD2 bench. Active optics are in orange. Blue
    disks mark focal planes and green disks pupil
    planes.}
  \label{fig:thd_bench}
\end{figure*}

We probe the coronagraphic performance of the component described in
section~\ref{sec:manufacturing} on the THD2
bench that has already been described in the
literature~\citep{baudoz18a,baudoz18b,patru18,potier18,singh19}.
Figure\,\ref{fig:thd_bench} shows the layout of the bench.
In
this paper, we removed the atmospheric residual wheel from the beam so
that we simulate a space-like telescope with an unobscured pupil
of~$8.23$\,mm diameter. A linear polarizer is
set in front of the Andor NEO detector. A flat mirror replaces the
deformable mirror~DM2. The wrapped vortex phase mask is set in focal
plane~$3$ (blue disk). The low-order wavefront sensor
\citep[LOWFS,][]{singh14} is used to stabilize the centering of the
beam onto the phase mask. Deformable mirrors, namely~DM3 set in a pupil
plane and~DM1 at~$26.9\,$cm from the pupil plane, are used to
control both phase and amplitude aberrations of the wavefront and
thereby minimize the speckle intensity in a~$360\,^{\circ}$
field of view around the optical axis (i.e., the star center). The
self-coherent camera is used for focal plane wavefront sensing and to
control the deformable mirrors
\citep{galicher10,mazoyer13a,mazoyer13b,mazoyer14a}.
Several sources of the injection system can light up the bench: laser
diodes ($638.6\pm2.3\,$nm, $704.5\pm2.1\,$nm and, $783\pm2.3\,$nm) or
the supercontinuum source associated with spectral filters with
a~$\sim10\,$nm bandwidth. The flux and the spectrum of the light that
enters the~THD2 bench are continuously measured so that the photometry
can easily be calibrated in the recorded images.

\subsection{Transmission of an off-axis source}
\label{subsec:exp_offpsf}
\begin{figure*}[!ht]
  \centering
  \includegraphics[width=\textwidth]{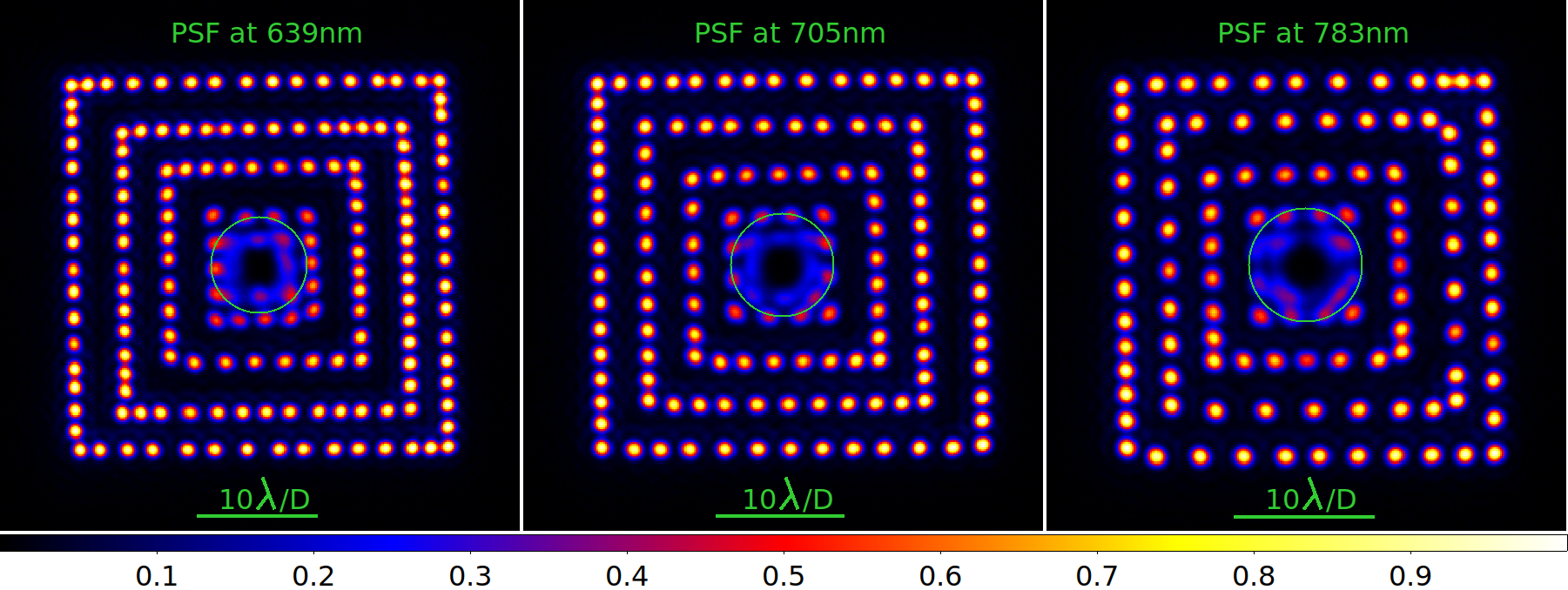}
  \caption{Laboratory measurement: Superposition of the images of
    off-axis sources at $639\,$nm (left), $705\,$nm (middle), and
    $783\,$nm (right). The color bar gives the transmission through the
    wrapped vortex coronagraph. Circles have a radius
    of~$4\,\lambda/D$ from the optical axis.}
  \label{fig:exp_offaxis_images}
\end{figure*}

\begin{figure*}[!ht]
  \centering
  \includegraphics[width=6cm]{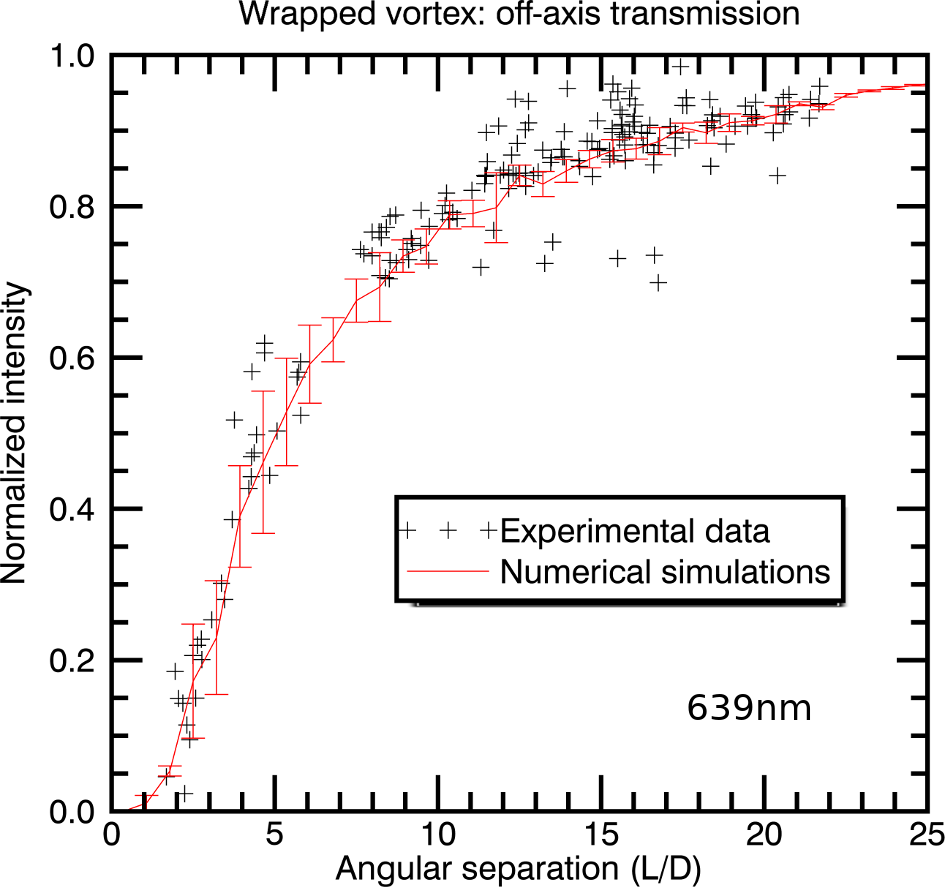}
  \includegraphics[width=6cm]{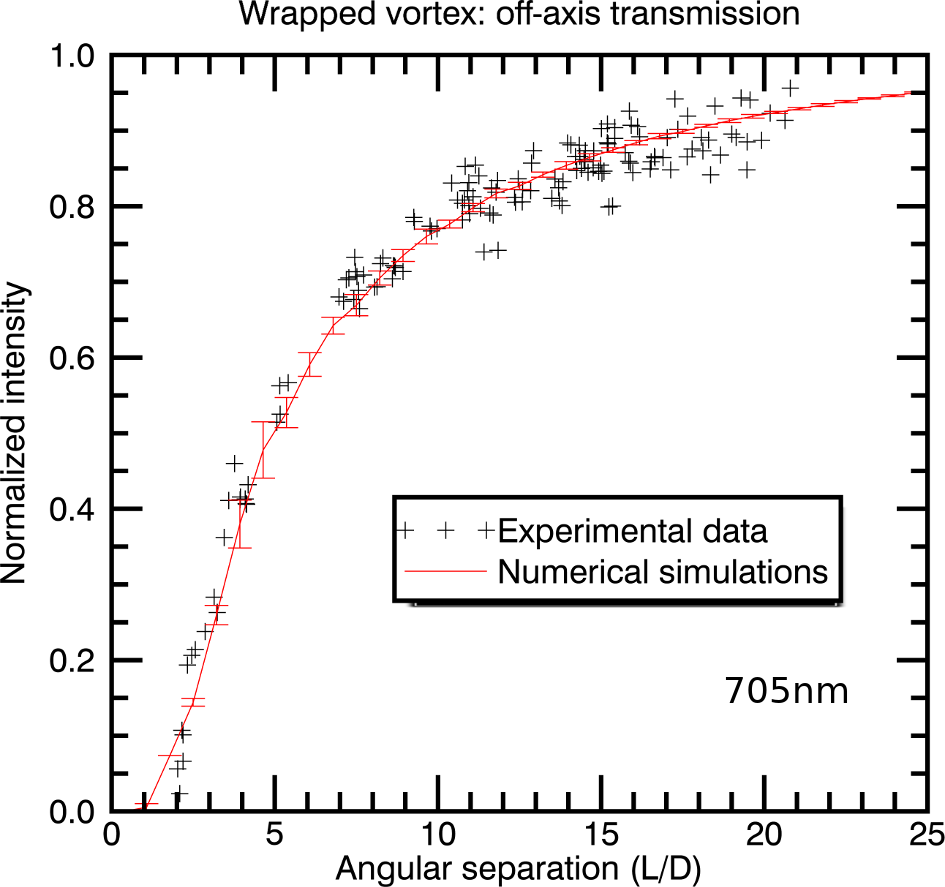}
  \includegraphics[width=6cm]{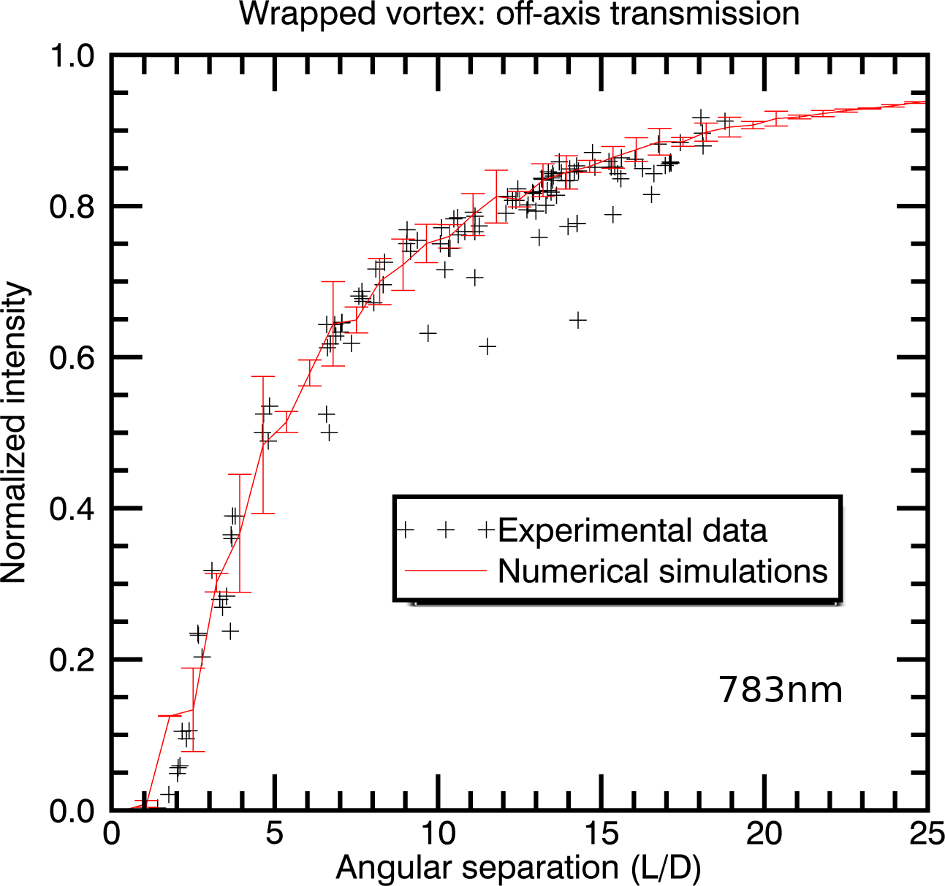}
  \caption{Laboratory measurement: Transmission of an off-axis source
    through the wrapped vortex coronagraph against the angular
    separation at three wavelengths: $639\,$nm (left), $705\,$nm
    (middle), and $783\,$nm (right). Predictions from numerical
    simulations are plotted in red.}
  \label{fig:exp_offaxis_curves}
\end{figure*}

In this section, we study the transmission of the coronagraph for an
off-axis source like an exoplanet, as done by numerical
simulations in section\,\ref{subsec:wrapped_offpsf}.
We record off-axis images at the three laser diode wavelengths
moving the beam in the field thanks to the tip-tilt mirror (dark brown
in figure~\ref{fig:thd_bench}). The Lyot
stop diameter is~$8.00\,$mm ($97\,\%$ of the entrance pupil
diameter). At maximum, we can move the source to~$\sim22\,\lambda/D$,
$\sim21\,\lambda/D$, and~$\sim19\,\lambda/D$ from the optical axis
at $639\,$nm, $705\,$nm, and $783\,$nm respectively. From numerical
simulations, we find the transmission of the coronagraph at these
separations is theoretically $\sim93.4\,\%$, $\sim92.9\,\%,$ and,
$90.5\,\%$ respectively.
Therefore, for each wavelength, we select the five farthest images. We
calculate the average~$s$ of their separations to the optical axis
(the five separations are within~$1\,\lambda/D$ of each
other). We then normalize all the images so that the transmission
at~$s$ is equal to the transmission found by numerical simulations in
section\,\ref{subsec:wrapped_offpsf}. Doing so, we suppose the
transmission of the fused silica that the mask is made of
is~$100\,\%$. This is a reasonable assumption because the coating is of
high quality (section\,\ref{subsec:spec}).
Figure\,\ref{fig:exp_offaxis_images} shows the superposition of the
normalized images of all the off-axis sources for the three
wavelengths.
The experimental transmissions of the wrapped vortex coronagraph
(i.e. the maximum of each off-axis image) are then plotted as a
function of the angular separation in
figure\,\ref{fig:exp_offaxis_curves}.
The curves obtained by numerical simulations are overplotted in
red. Curves and experimental data are in good agreement at
the positions used for the normalization of the data (the farthest
separations from the star). They are also consistent at all
angular separations, validating the results of our numerical
simulations. In conclusion, we find that the inner working angle at which
the transmission of an off-axis source reaches~$50\,\%$ is
between~$4\,\lambda/D$ and $6\,\lambda/D$ for the wrapped vortex phase
mask.

\subsection{Coronagraphic attenuation}
\label{subsec:exp_att}
In the second laboratory experiment, we measure the coronagraphic
attenuation of the star image as done by numerical simulations in
section~\ref{subsec:att}.
The Lyot stop diameter is~$7.90\,$mm ($96\,\%$ of the entrance pupil
diameter). We first minimize the speckle intensity at~$783\,$nm (laser
diode) in a $27\,\lambda/D\ \times\ 27\,\lambda/D$ square centered on
the optical axis: we measure the electric field using the
self-coherent camera and we use both deformable mirrors~DM1
and~DM3 to compensate for phase and amplitude aberrations. The
deformable mirror shapes are then fixed and we light up 
the system at different wavelengths. Finally, we record coronagraphic
images and noncoronagraphic images for photometry calibration.
Noncoronagraphic images are recorded in the presence
of the wrapped vortex phase mask but at $\sim20\,\lambda/D$
from the center of the mask. We show in
section~\ref{subsec:exp_offpsf} that the transmission of the
coronagraph at such separations is below~$95\,\%$. Therefore, we use
the transmission curve calculated by numerical simulations to
calibrate the noncoronagraphic images.

\begin{figure*}[!ht]
  \centering
  \includegraphics[width=.7\textwidth]{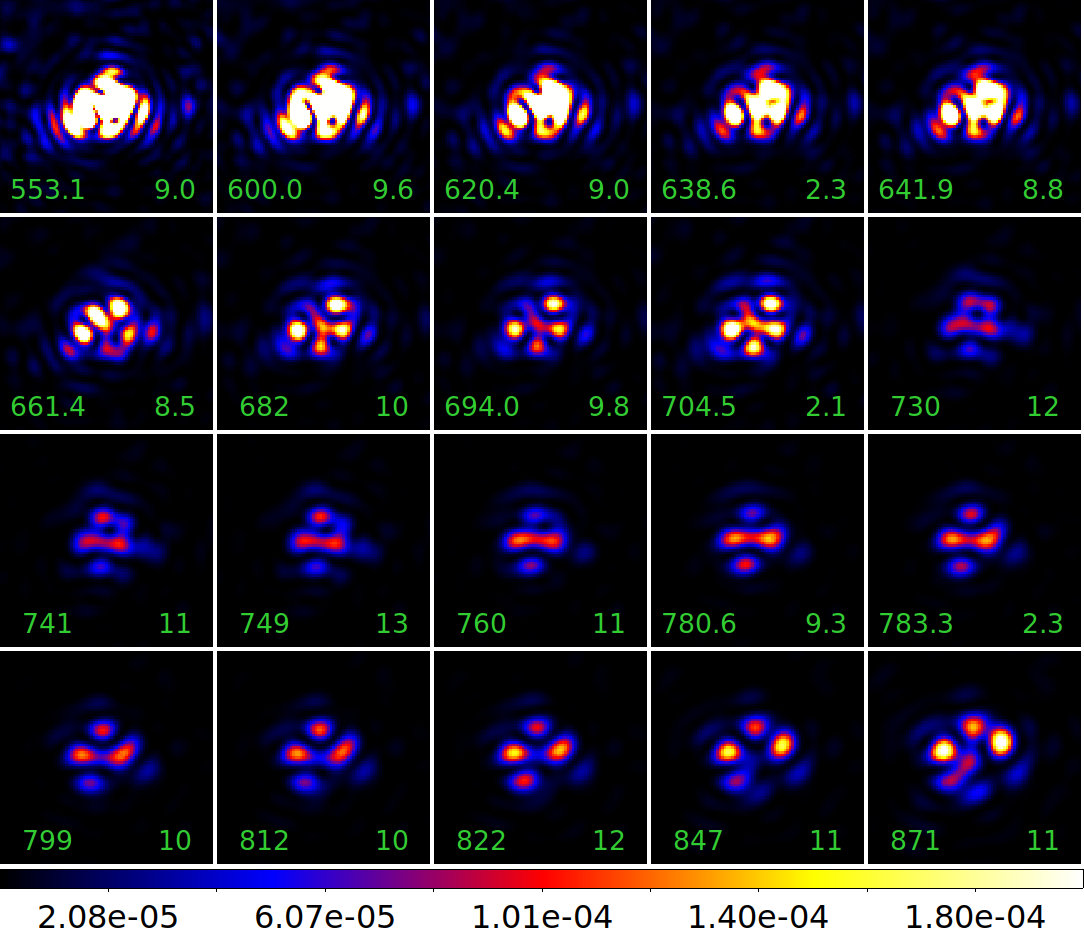}
  \caption{Laboratory measurement: Coronagraphic images recorded at
    different wavelengths (central wavelength in the bottom left and
    spectral bandwidth in the bottom right of each image). Speckles
    are minimized using the laser diode at~$783\,$nm. The image size
    is~$12\,\lambda/D\times12\,\lambda/D$ at~$704.5\,$nm. The color bar
    gives the intensity normalized by the maximum of the
    noncoronagraphic image.}
  \label{fig:exp_centre_image_783nm}
\end{figure*}

\begin{figure}[!ht]
  \centering
  \includegraphics[width=.45\textwidth]{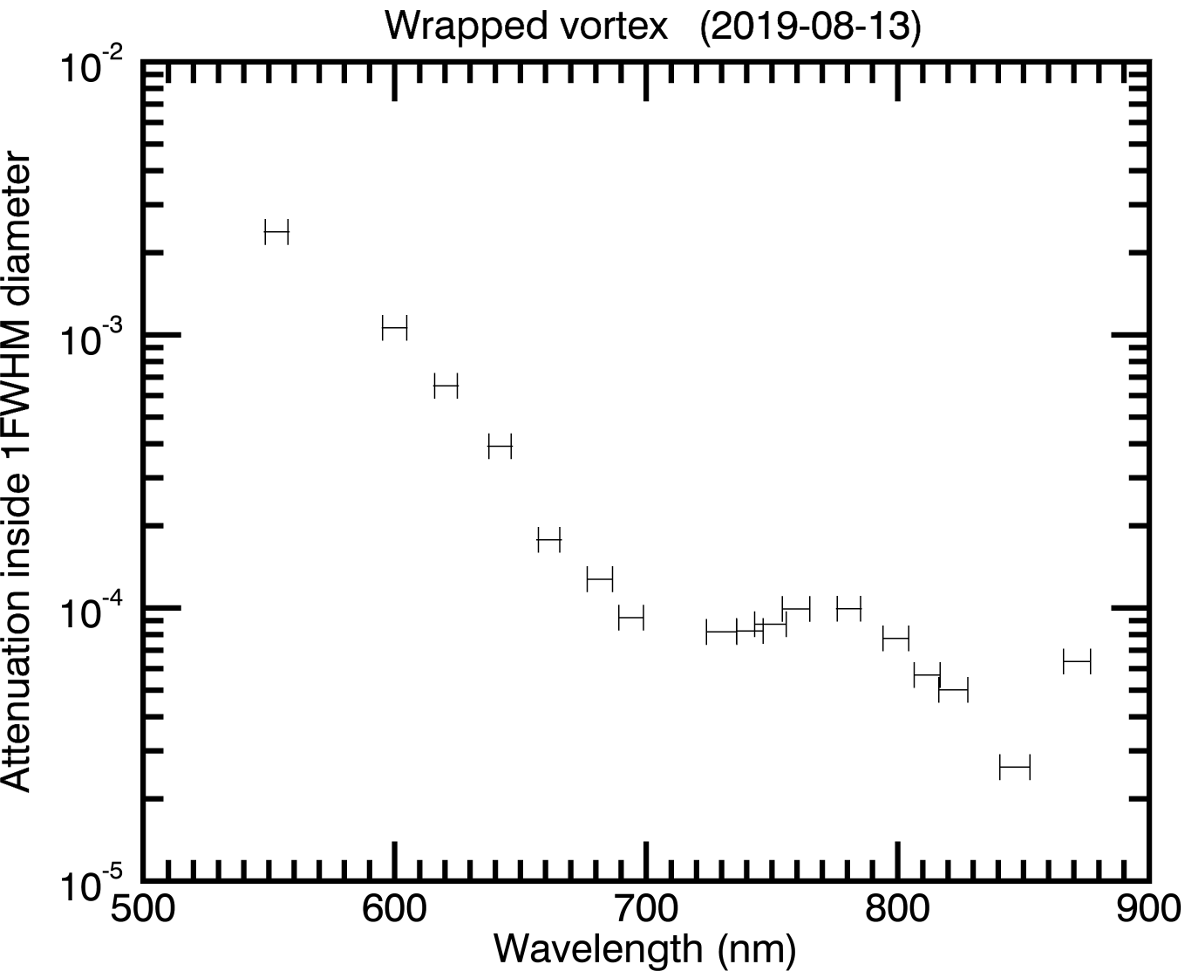}
  \caption{Laboratory measurement: Coronagraphic attenuation~$A$
    integrated inside a disk of~$1\,\lambda/D$-diameter centered on
    the optical axis. The horizontal excursions give the filter
    bandwidth.}
  \label{fig:exp_att}
\end{figure}

Figure\,\ref{fig:exp_centre_image_783nm} shows the coronagraphic
images. The central wavelength and the spectral bandwidth of the
filter (or the laser diode) are written in the bottom left and right
of each image respectively.
The color bar gives the intensity normalized by the maximum of the
noncoronagraphic image. The images show the same speckle pattern
between~$\sim680\,$ and~$870\,$nm. Below~$680\,$nm, images slightly
evolve and the attenuation degrades. This is also visible in
figure~\,\ref{fig:exp_att}, which plots the coronagraphic
attenuation~$A$ defined in section~\ref{subsec:att} as a function of
wavelength. The horizontal excursions give the filter bandwidth.
The attenuation is better than~$10^{-4}$ at least between~$680\,$nm
and~$870\,$mn corresponding to a~$>25\,\%$ bandpass. We could not
measure the performance at wavelengths larger than~$870\,$nm which is
the current cut-off of the injection system. This result is consistent
with the numerical simulations that predict an attenuation better
than~$10^{-4}$ over a~$29\,\%$ bandpass
(figure~\,\ref{fig:attenwrappedvortex}) and with the expected central
wavelength of~$\sim770\,$nm
(section~\ref{subsec:delivered_component}). The point at~$850\,$nm is 
lower than its neighbors although there are bright speckles in the
image (figure\,\ref{fig:exp_centre_image_783nm}). The speckles are
further away than~$0.5\,\lambda/D$ from the center because of spectral
dispersion.

\subsection{Coronagraphic dark holes}
The third experiment is an analysis of the ability to reach very high
contrast at several wavelengths using a wrapped vortex phase mask.
In an optimal protocol, one would optimize the~DM1 and~DM3 deformable
mirror shapes at a central wavelength to create a dark hole (i.e., region of
high contrast in the coronagraphic image). One would then light the
bench up with several sources to record dark hole images at all
wavelengths. Using the wrapped vortex phase mask and another broadband
coronagraph -- vector vortex plus polarizers -- we recently noticed
that there is a slow evolution of the speckle pattern at mid-spatial
frequencies. Therefore, there are small chromatic aberrations on the
current bench and we are still investigating the reasons for this
evolution.

Because of these aberrations, we have to slightly modify the
deformable mirror shapes to minimize the speckle intensity at each
wavelength. Hence, in this section, for each laser diode
($638.6\pm2.3\,$nm, $704.5\pm2.1\,$nm and, $783\pm2.3\,$nm), we
independently measure the electric field associated to the stellar
speckles using the self-coherent camera, and we minimize the star
intensity inside a dark hole. We record coronagraphic images and then
noncoronagraphic images moving the source image
to~$\sim20\,\lambda/D$ from the center of the phase mask. As in
section~\ref{subsec:exp_att}, we correct the photometry of the
noncoronagraphic images from the transmission of the coronagraph that
is not exactly~$100\,\%$ at these separations. Finally, we normalize
the coronagraphic images by the maximum of the noncoronagraphic image
and correct each pixel by the coronagraph transmission using the
curve obtained in numerical simulations
(figure~\ref{fig:exp_offaxis_curves}).

Results for a $360^{\circ}$ full dark hole are presented in
section~\ref{subsubsec:FDH}. Deeper contrast levels are then obtained
when reducing the dark hole size as shown in
section~\ref{subsubsec:HDH}. For these measurements, the Lyot stop
diameter is~$6.50\,$mm ($79\,\%$ of the entrance pupil diameter).

\subsubsection{Full dark hole of $360^{\circ}$}
\label{subsubsec:FDH}
\begin{figure*}[!ht]
  \centering
  \includegraphics[width=\textwidth]{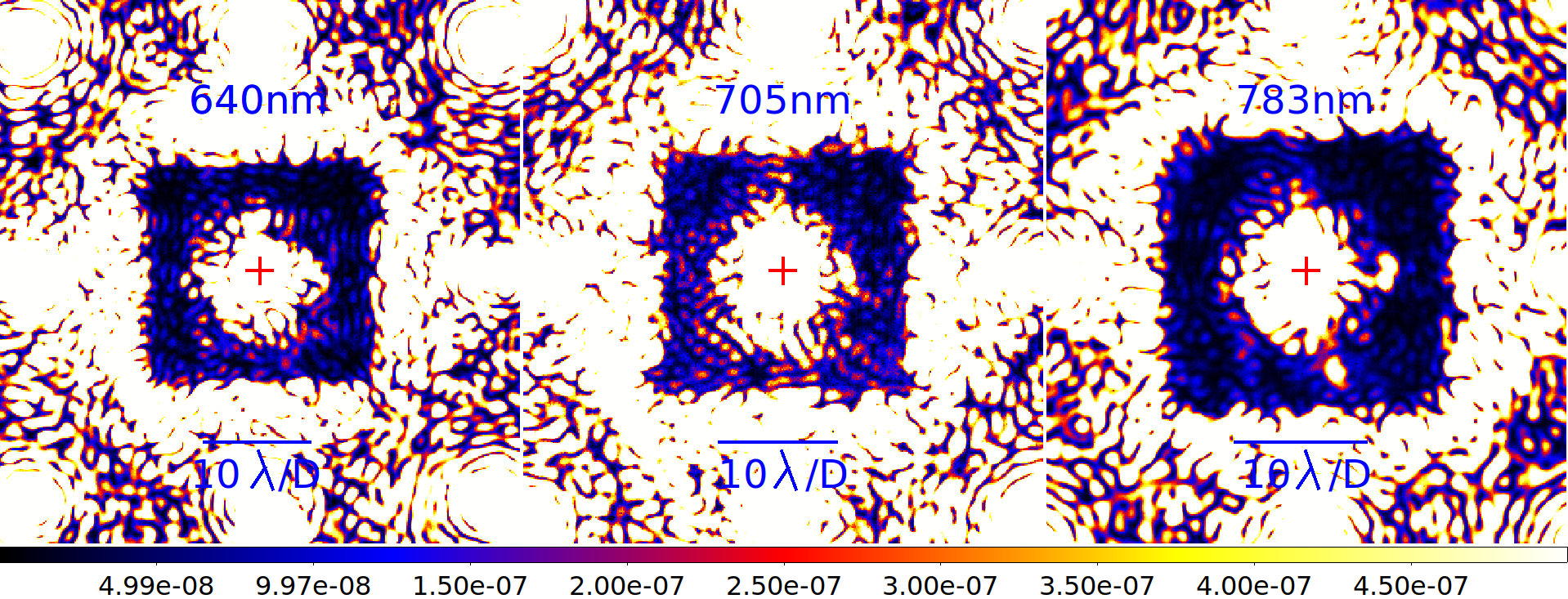}
  \caption{Laboratory measurement: Coronagraphic images obtained
    at~$639\,$nm (left), $705\,$nm (center), and $783\,$nm (right) with
    a~$27\,\lambda/D\ \times\ 27\,\lambda/D$ dark hole. The center of
    the star image is at the center of the field of view (red
    cross). Each image is corrected by the coronagraph
    transmission. The color bar gives the intensity normalized by the
    maximum of the noncoronagraphic image.}
  \label{fig:exp_FDH_im}
\end{figure*}

\begin{figure}[!ht]
  \centering
  \includegraphics[width=7cm]{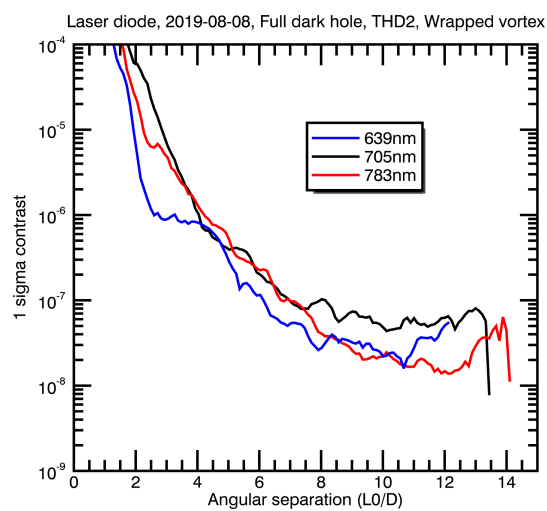}
  \caption{Laboratory measurement: $1\,\sigma$ contrast curves
    associated with the 
    full dark hole coronagraphic images of 
    figure\,\ref{fig:exp_FDH_im} at~$639\,$nm (blue), $705\,$nm
    (black), and $783\,$nm (red). The horizontal axis gives the angular
    separation from the star in~$\lambda_0/D$
    with~$\lambda_0=705\,$nm. These curves are corrected by the
    coronagraph transmission of an off-axis source
    (figure~\ref{fig:exp_offaxis_curves}).}
  \label{fig:exp_FDH_curve}
\end{figure}

We first minimize the speckle intensity in a square dark
hole of~$27\,\lambda/D\ \times\ 27\,\lambda/D$ centered on the
optical axis. The coronagraphic images are presented in
figure\,\ref{fig:exp_FDH_im}.
The color bar gives the intensity normalized by the maximum of the
noncoronagraphic image. As coronagraphic images are corrected by the
coronagraph transmission, the color bar indicates the ratio between the
flux of the exoplanet that can be detected and the star. In
figure\,\ref{fig:exp_FDH_curve}, we plot the~$1\,\sigma$ contrast curve
that is the azimuthal standard deviation in annuli
of~$0.5\,\lambda/D$ for each image of figure~\ref{fig:exp_FDH_im}.
The horizontal axis gives the angular separation from the star
in~$\lambda_0/D$ with~$\lambda_0=705\,$nm.
Inside the dark holes, regions of interest where we would look for
exoplanet images, the~$3\,\sigma$ contrast level is better
than~$3.10^{-7}$ over a~$20\,\%$ spectral bandpass at~$5-7\,\lambda_0/D$
or more from the star.

\subsubsection{Half-field-of-view dark hole}
\label{subsubsec:HDH}
In order to reach deeper contrast levels, we reduce the dark hole to a
region going from~$1\,\lambda/D$ to~$13.5\,\lambda/D$ in one direction
and from~$-13.5\,\lambda/D$ to~$13.5\,\lambda/D$ in the other direction.
\begin{figure*}[!ht]
  \centering
  \includegraphics[width=\textwidth]{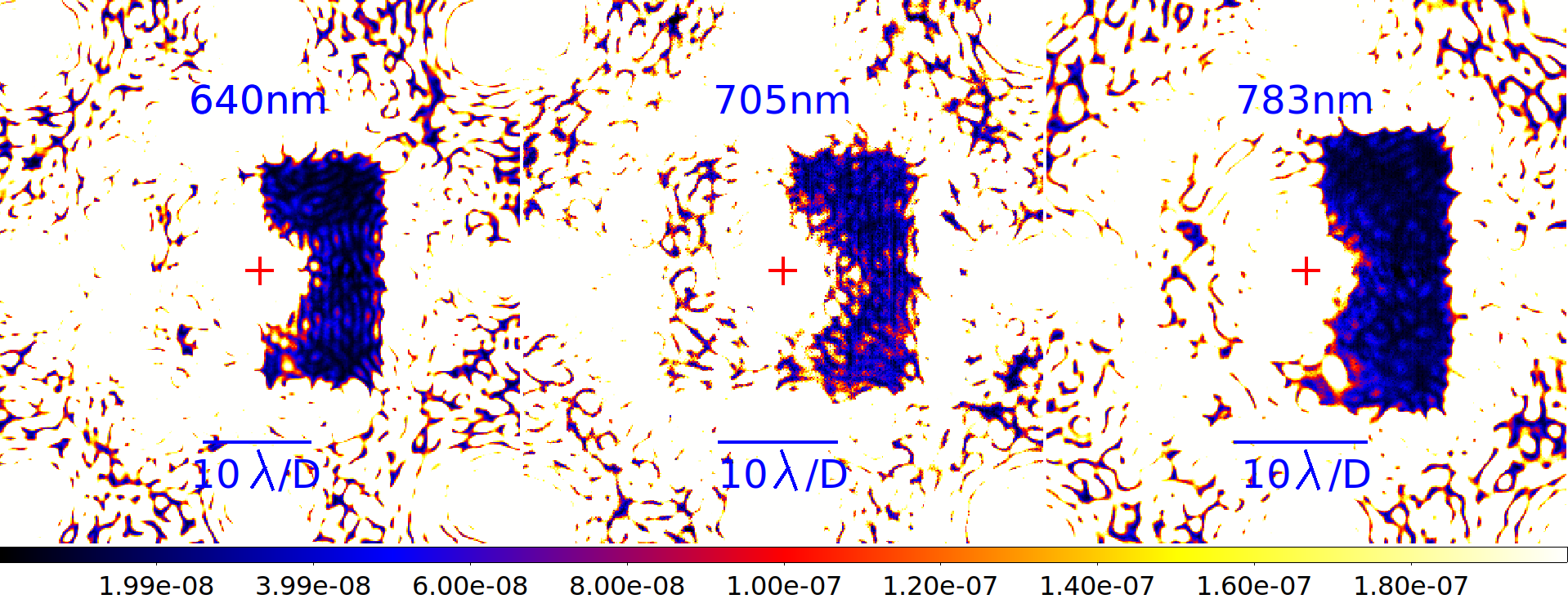}
  \caption{Laboratory measurement: Coronagraphic images obtained
    at~$639\,$nm (left), $705\,$nm (center), and $783\,$nm (right) with
    a half dark hole. The center of the star image is at the center of
    the field of view (red cross). Each image is corrected by the
    coronagraph transmission. The color bar gives the intensity
    normalized by the maximum of the noncoronagraphic image.}
  \label{fig:exp_HDH_im}
\end{figure*}
 Images at the three wavelengths are shown in
figure~\ref{fig:exp_HDH_im} and the~$1\,\sigma$ contrast curves are
plotted in figure~\ref{fig:exp_HDH_curve}. The contrast is plotted
against the angular separation expressed in~$\lambda_0/D$
with~$\lambda_0=705\,$nm.
\begin{figure}[!ht]
  \centering
  \includegraphics[width=7cm]{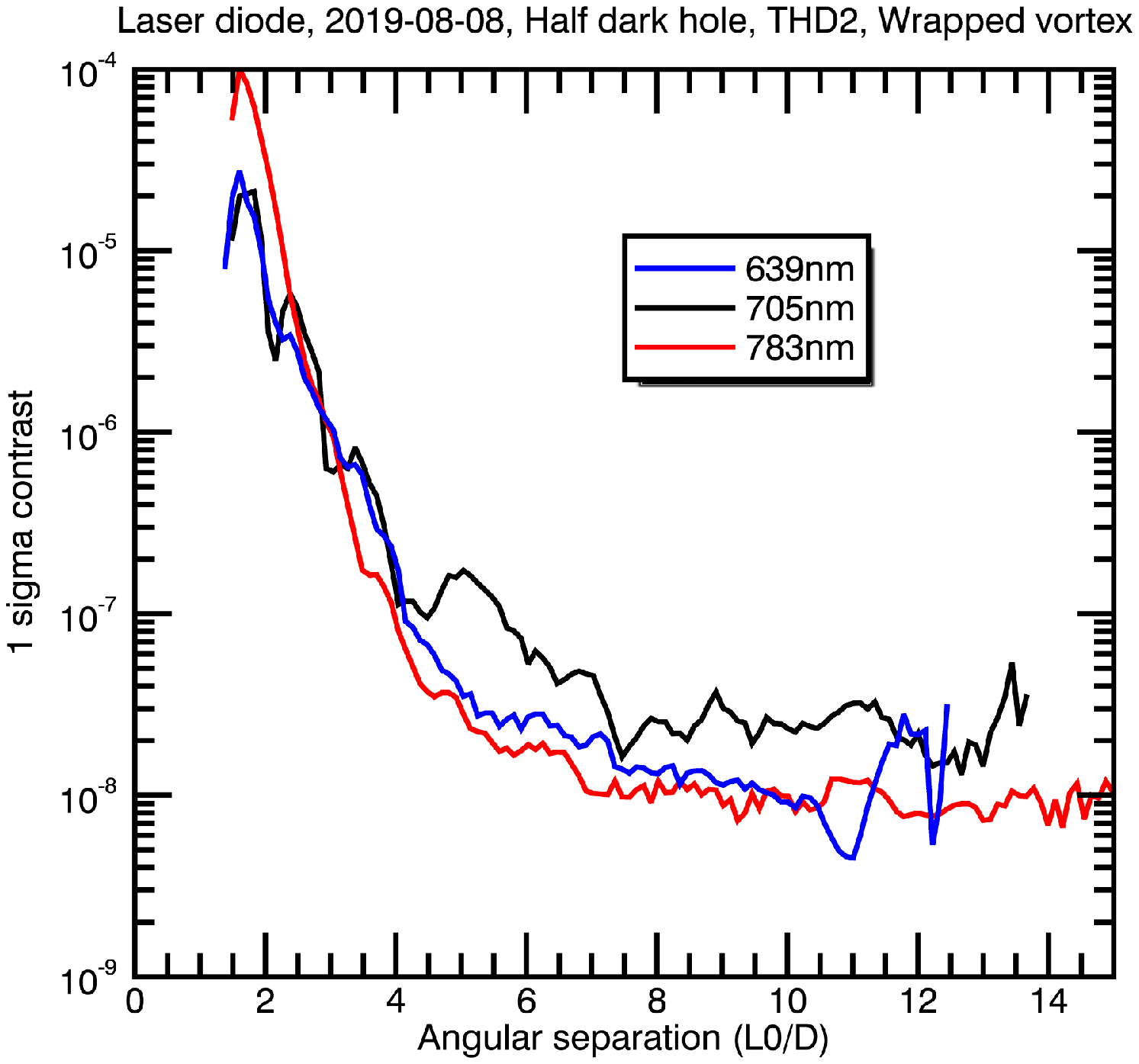}
  \caption{Laboratory measurement: $1\,\sigma$ contrast curves
    associated with the half dark hole coronagraphic images of 
    figure\,\ref{fig:exp_HDH_im} at~$639\,$nm (blue), $705\,$nm
    (black), and $783\,$nm (red). The horizontal axis gives the angular
    separation from the star in~$\lambda_0/D$
    with~$\lambda_0=705\,$nm. These curves are corrected by the
    coronagraph transmission of an off-axis source
    (figure~\ref{fig:exp_offaxis_curves}).} 
  \label{fig:exp_HDH_curve}
\end{figure}
At~$705\,$nm, the performance is worse than at the two other
wavelengths because of detector noise that is visible in the image. In
the~$705\,$nm image, an exoplanet~$10^7$ times fainter than its star
could be detected at~$3\,\sigma$ in the raw image (prior to any
post-processing of the data). At~$639\,$nm and~$783\,$nm, an exoplanet
up to~$\sim3.10^7$ times fainter than its star could be detected
at~$3\,\sigma$ between~$7\,\lambda_0/$
and~$12-14\,\lambda_0/D$. Numerical simulations of such a dark hole
using the wrapped vortex phase mask on the THD2 bench predict similar
performance with a~$3\,\sigma$ detection limit at~$\sim3.10^{-8}$
between~$6\,\lambda_0/D$ and~$13\,\lambda_0/D$. In conclusion, the
laboratory performance demonstrates the efficiency of the wrapped
vortex coronagraph over broadband of at least~$20\,\%$.

\section{Conclusions}
We used a mathematical approach to derive a family of 2D azimuthal
phase functions for broadband coronagraphy. One of the functions is a
wrapped vortex that we studied by numerical simulations. We predicted
very efficient coronagraphic attenuation over a~$30\,\%$ spectral
bandpass with very high transmission of the exoplanet light and
a~$360^{\circ}$ field of view. Taking advantage of a new technology
for etching continuous 2D functions, we manufactured a prototype of
the wrapped vortex. We demonstrated in the laboratory that it attenuates
the starlight by a factor of~$10^4$ or more over a~$>25\,\%$
bandpass. Finally, we obtained dark hole images with a~$3\,\sigma$
detection limit at~$\sim3.10^{-8}$ at~$639\,$nm and $783\,$nm. Aside
from very efficient broadband coronagraphic performance, the wrapped
vortex requires only one scalar focal plane mask followed by a
classical binary Lyot stop. Such a simple design is more attractive
than other coronagraphs, which need several optics to be aligned and
stabilized like the vector vortex that requires the use of polarizers,
or multi-stage devices. Moreover, as a scalar phase mask no
polarization filtering is needed and~$\sim100\,\%$ of the putative
exoplanet light is transmitted towards the science detector. One
  drawback of the wrapped vortex is the inner working angle that
  is~$5\,\lambda/D$. Smaller inner working angles can be achieved with
  single phase masks such as the FQPM or SLPM at the expense of a
  narrower spectral band. In future
work, families of radially and azimuthally varying wrapped
functions will be investigated so that they can be used with obscured
pupils.

\section{Acknowledgment}
French co-authors are supported by ANR-14-CE33-0018.
Part of this work was supported by the Physics department of
the University of Paris.

\bibliographystyle{aa}   
\bibliography{/home/raphael/biblio/bibtexbase/database.bib}   

\begin{thebibliography}{43}
\expandafter\ifx\csname natexlab\endcsname\relax\def\natexlab#1{#1}\fi

\bibitem[{{Adibekyan}(2019)}]{Adibekyan19}
{Adibekyan}, V. 2019, Geosciences, 9, 105

\bibitem[{{Alonso-Floriano} {et~al.}(2019){Alonso-Floriano},
  {S{\'a}nchez-L{\'o}pez}, {Snellen}, {L{\'o}pez-Puertas}, {Nagel}, {Amado},
  {Bauer}, {Caballero}, {Czesla}, \& {Nortmann}}]{alonso19}
{Alonso-Floriano}, F.~J., {S{\'a}nchez-L{\'o}pez}, A., {Snellen}, I.~A.~G.,
  {et~al.} 2019, Astronomy and Astrophysics, 621, A74

\bibitem[{Baba \& Murakami(2003)}]{baba03}
Baba, N. \& Murakami, N. 2003, Publications of the Astronomical Society of the
  Pacific, 115, 1363

\bibitem[{{Baudoz} {et~al.}(2018{\natexlab{a}}){Baudoz}, {Galicher}, {Patru},
  {Dupuis}, \& {Thijs}}]{baudoz18a}
{Baudoz}, P., {Galicher}, R., {Patru}, F., {Dupuis}, O., \& {Thijs}, S.
  2018{\natexlab{a}}, arXiv e-prints

\bibitem[{{Baudoz} {et~al.}(2018{\natexlab{b}}){Baudoz}, {Galicher}, {Potier},
  {Dupuis}, {Thijs}, \& {Patru}}]{baudoz18b}
{Baudoz}, P., {Galicher}, R., {Potier}, A., {et~al.} 2018{\natexlab{b}}, in
  Society of Photo-Optical Instrumentation Engineers (SPIE) Conference Series,
  Vol. 10706, Advances in Optical and Mechanical Technologies for Telescopes
  and Instrumentation III, 107062O

\bibitem[{{Beuzit} {et~al.}(2019){Beuzit}, {Vigan}, {Mouillet}, {Dohlen},
  {Gratton}, {Boccaletti}, {Sauvage}, {Schmid}, {Langlois}, {Petit},
  {Baruffolo}, {Feldt}, {Milli}, {Wahhaj}, {Abe}, {Anselmi}, {Antichi},
  {Barette}, {Baudrand}, {Baudoz}, {Bazzon}, {Bernardi}, {Blanchard}, {Brast},
  {Bruno}, {Buey}, {Carbillet}, {Carle}, {Cascone}, {Chapron}, {Charton},
  {Chauvin}, {Claudi}, {Costille}, {De Caprio}, {de Boer}, {Delboulb{\'e}},
  {Desidera}, {Dominik}, {Downing}, {Dupuis}, {Fabron}, {Fantinel}, {Farisato},
  {Feautrier}, {Fedrigo}, {Fusco}, {Gigan}, {Ginski}, {Girard}, {Giro},
  {Gisler}, {Gluck}, {Gry}, {Henning}, {Hubin}, {Hugot}, {Incorvaia}, {Jaquet},
  {Kasper}, {Lagadec}, {Lagrange}, {Le Coroller}, {Le Mignant}, {Le Ruyet},
  {Lessio}, {Lizon}, {Llored}, {Lundin}, {Madec}, {Magnard}, {Marteaud},
  {Martinez}, {Maurel}, {M{\'e}nard}, {Mesa}, {M{\"o}ller-Nilsson}, {Moulin},
  {Moutou}, {Orign{\'e}}, {Parisot}, {Pavlov}, {Perret}, {Pragt}, {Puget},
  {Rabou}, {Ramos}, {Reess}, {Rigal}, {Rochat}, {Roelfsema}, {Rousset}, {Roux},
  {Saisse}, {Salasnich}, {Santambrogio}, {Scuderi}, {Segransan}, {Sevin},
  {Siebenmorgen}, {Soenke}, {Stadler}, {Suarez}, {Tiph{\`e}ne}, {Turatto},
  {Udry}, {Vakili}, {Waters}, {Weber}, {Wildi}, {Zins}, \& {Zurlo}}]{beuzit19}
{Beuzit}, J.~L., {Vigan}, A., {Mouillet}, D., {et~al.} 2019, Astronomy and
  Astrophysics paper, 631, A155

\bibitem[{{Bonafous} {et~al.}(2016){Bonafous}, {Galicher}, {Baudoz}, {Firminy},
  \& {Boussaha}}]{bonafous16}
{Bonafous}, M., {Galicher}, R., {Baudoz}, P., {Firminy}, J., \& {Boussaha}, F.
  2016, in Society of Photo-Optical Instrumentation Engineers (SPIE) Conference
  Series, Vol. 9912, Proceedings of the SPIE, 99126J

\bibitem[{{Cady} {et~al.}(2017){Cady}, {Balasubramanian}, {Gersh-Range},
  {Kasdin}, {Kern}, {Lam}, {Mejia Prada}, {Moody}, {Patterson}, {Poberezhskiy},
  {Riggs}, {Seo}, {Shi}, {Tang}, {Trauger}, {Zhou}, \& {Zimmerman}}]{cady17}
{Cady}, E., {Balasubramanian}, K., {Gersh-Range}, J., {et~al.} 2017, in Society
  of Photo-Optical Instrumentation Engineers (SPIE) Conference Series, Vol.
  10400, Proceedings of the SPIE, 104000E

\bibitem[{{Close} {et~al.}(2018){Close}, {Males}, {Morzinski}, {Esposito},
  {Riccardi}, {Briguglio}, {Follette}, {Wu}, {Pinna}, {Puglisi}, {Xompero},
  {Quiros}, \& {Hinz}}]{close18}
{Close}, L.~M., {Males}, J.~R., {Morzinski}, K.~M., {et~al.} 2018, in Society
  of Photo-Optical Instrumentation Engineers (SPIE) Conference Series, Vol.
  10703, Proceedings of the SPIE, 107030L

\bibitem[{{D'Angelo} \& {Lissauer}(2018)}]{dangelo18}
{D'Angelo}, G. \& {Lissauer}, J.~J. 2018, {Formation of Giant Planets}, 140

\bibitem[{{Espinoza} {et~al.}(2019){Espinoza}, {Rackham}, {Jord{\'a}n}, {Apai},
  {L{\'o}pez-Morales}, {Osip}, {Grimm}, {Hoeijmakers}, {Wilson}, \&
  {Bixel}}]{espinoza19}
{Espinoza}, N., {Rackham}, B.~V., {Jord{\'a}n}, A., {et~al.} 2019, Monthly
  Notices of the Royal Astronomical Society, 482, 2065

\bibitem[{Galicher {et~al.}(2011)Galicher, Baudoz, \& Baudrand}]{galicher11}
Galicher, R., Baudoz, P., \& Baudrand, J. 2011, Astronomy and Astrophysics,
  530, A43+

\bibitem[{{Galicher} {et~al.}(2010){Galicher}, {Baudoz}, {Rousset}, {Totems},
  \& {Mas}}]{galicher10}
{Galicher}, R., {Baudoz}, P., {Rousset}, G., {Totems}, J., \& {Mas}, M. 2010,
  Astronomy and Astrophysics, 509, A31+

\bibitem[{{Gravity Collaboration} {et~al.}(2019){Gravity Collaboration},
  {Lacour}, {Nowak}, {Wang}, {Pfuhl}, {Eisenhauer}, {Abuter}, {Amorim},
  {Anugu}, {Benisty}, {Berger}, {Beust}, {Blind}, {Bonnefoy}, {Bonnet},
  {Bourget}, {Brandner}, {Buron}, {Collin}, {Charnay}, {Chapron}, {Cl{\'e}net},
  {Coud{\'e} Du Foresto}, {de Zeeuw}, {Deen}, {Dembet}, {Dexter}, {Duvert},
  {Eckart}, {F{\"o}rster Schreiber}, {F{\'e}dou}, {Garcia}, {Garcia Lopez},
  {Gao}, {Gendron}, {Genzel}, {Gillessen}, {Gordo}, {Greenbaum}, {Habibi},
  {Haubois}, {Hau{\ss}mann}, {Henning}, {Hippler}, {Horrobin}, {Hubert},
  {Jimenez Rosales}, {Jocou}, {Kendrew}, {Kervella}, {Kolb}, {Lagrange},
  {Lapeyr{\`e}re}, {Le Bouquin}, {L{\'e}na}, {Lippa}, {Lenzen}, {Maire},
  {Molli{\`e}re}, {Ott}, {Paumard}, {Perraut}, {Perrin}, {Pueyo}, {Rabien},
  {Ram{\'\i}rez}, {Rau}, {Rodr{\'\i}guez-Coira}, {Rousset}, {Sanchez-Bermudez},
  {Scheithauer}, {Schuhler}, {Straub}, {Straubmeier}, {Sturm}, {Tacconi},
  {Vincent}, {van Dishoeck}, {von Fellenberg}, {Wank}, {Waisberg}, {Widmann},
  {Wieprecht}, {Wiest}, {Wiezorrek}, {Woillez}, {Yazici}, {Ziegler}, \&
  {Zins}}]{lacour19}
{Gravity Collaboration}, {Lacour}, S., {Nowak}, M., {et~al.} 2019, Astronomy
  and Astrophysics paper, 623, L11

\bibitem[{{Hou} {et~al.}(2014){Hou}, {Cao}, {Zhu}, \& {Ma}}]{hou14}
{Hou}, F., {Cao}, Q., {Zhu}, M., \& {Ma}, O. 2014, Optics Express, 22, 1884

\bibitem[{{Izidoro} \& {Raymond}(2018)}]{Izidoro18}
{Izidoro}, A. \& {Raymond}, S.~N. 2018, {Formation of Terrestrial Planets}, 142

\bibitem[{{Jovanovic} {et~al.}(2018){Jovanovic}, {Absil}, {Baudoz}, {Beaulieu},
  {Bottom}, {Cady}, {Carlomagno}, {Carlotti}, {Doelman}, {Fogarty}, {Galicher},
  {Guyon}, {Haffert}, {Huby}, {Jewell}, {Keller}, {Kenworthy}, {Knight},
  {K{\"u}hn}, {Miller}, {Mazoyer}, {N'Diaye}, {Por}, {Pueyo}, {Riggs}, {Ruane},
  {Sirbu}, {Snik}, {Wallace}, {Wilby}, \& {Ygouf}}]{jovanovic18}
{Jovanovic}, N., {Absil}, O., {Baudoz}, P., {et~al.} 2018, in Society of
  Photo-Optical Instrumentation Engineers (SPIE) Conference Series, Vol. 10703,
  Proceedings of the SPIE, 107031U

\bibitem[{{Lozi} {et~al.}(2018){Lozi}, {Guyon}, {Jovanovic}, {Goebel},
  {Pathak}, {Skaf}, {Sahoo}, {Norris}, {Martinache}, {N'Diaye}, {Mazin},
  {Walter}, {Tuthill}, {Kudo}, {Kawahara}, {Kotani}, {Ireland}, {Cvetojevic},
  {Huby}, {Lacour}, {Vievard}, {Groff}, {Chilcote}, {Kasdin}, {Knight}, {Snik},
  {Doelman}, {Minowa}, {Clergeon}, {Takato}, {Tamura}, {Currie}, {Takami}, \&
  {Hayashi}}]{lozi18}
{Lozi}, J., {Guyon}, O., {Jovanovic}, N., {et~al.} 2018, in Society of
  Photo-Optical Instrumentation Engineers (SPIE) Conference Series, Vol. 10703,
  Proceedings of the SPIE, 1070359

\bibitem[{{Ma} {et~al.}(2012){Ma}, {Cao}, \& {Hou}}]{ma12}
{Ma}, O., {Cao}, Q., \& {Hou}, F. 2012, Optics Express, 20, 10933

\bibitem[{{Macintosh} {et~al.}(2014){Macintosh}, {Graham}, {Ingraham},
  {Konopacky}, {Marois}, {Perrin}, {Poyneer}, {Bauman}, {Barman}, {Burrows},
  {Cardwell}, {Chilcote}, {De Rosa}, {Dillon}, {Doyon}, {Dunn}, {Erikson},
  {Fitzgerald}, {Gavel}, {Goodsell}, {Hartung}, {Hibon}, {Kalas}, {Larkin},
  {Maire}, {Marchis}, {Marley}, {McBride}, {Millar-Blanchaer}, {Morzinski},
  {Norton}, {Oppenheimer}, {Palmer}, {Patience}, {Pueyo}, {Rantakyro},
  {Sadakuni}, {Saddlemyer}, {Savransky}, {Serio}, {Soummer},
  {Sivaramakrishnan}, {Song}, {Thomas}, {Wallace}, {Wiktorowicz}, \&
  {Wolff}}]{macintosh14}
{Macintosh}, B., {Graham}, J.~R., {Ingraham}, P., {et~al.} 2014, Proceedings of
  the National Academy of Science, 111, 12661

\bibitem[{{Marois} {et~al.}(2006){Marois}, {Lafreni{\`e}re}, {Doyon},
  {Macintosh}, \& {Nadeau}}]{marois06}
{Marois}, C., {Lafreni{\`e}re}, D., {Doyon}, R., {Macintosh}, B., \& {Nadeau},
  D. 2006, The Astrophysical Journal, 641, 556

\bibitem[{Mawet {et~al.}(2005)Mawet, Riaud, Absil, \& Surdej}]{Mawet05}
Mawet, D., Riaud, P., Absil, O., \& Surdej, J. 2005, The Astrophysical Journal,
  633, 1191

\bibitem[{{Mawet} {et~al.}(2009){Mawet}, {Serabyn}, {Liewer}, {Hanot},
  {McEldowney}, {Shemo}, \& {O'Brien}}]{mawet09}
{Mawet}, D., {Serabyn}, E., {Liewer}, K., {et~al.} 2009, Optics Express, 17,
  1902

\bibitem[{{Mawet} {et~al.}(2011){Mawet}, {Serabyn}, {Wallace}, \&
  {Pueyo}}]{mawet11}
{Mawet}, D., {Serabyn}, E., {Wallace}, J.~K., \& {Pueyo}, L. 2011, Optics
  Letters, 36, 1506

\bibitem[{{Mazoyer} {et~al.}(2013{\natexlab{a}}){Mazoyer}, {Baudoz},
  {Galicher}, {Mas}, \& {Rousset}}]{mazoyer13a}
{Mazoyer}, J., {Baudoz}, P., {Galicher}, R., {Mas}, M., \& {Rousset}, G.
  2013{\natexlab{a}}, Astronomy and Astrophysics, 557, A9

\bibitem[{{Mazoyer} {et~al.}(2013{\natexlab{b}}){Mazoyer}, {Baudoz},
  {Galicher}, \& {Rousset}}]{mazoyer13b}
{Mazoyer}, J., {Baudoz}, P., {Galicher}, R., \& {Rousset}, G.
  2013{\natexlab{b}}, in Society of Photo-Optical Instrumentation Engineers
  (SPIE) Conference Series, Vol. 8864, Society of Photo-Optical Instrumentation
  Engineers (SPIE) Conference Series

\bibitem[{{Mazoyer} {et~al.}(2014){Mazoyer}, {Baudoz}, {Galicher}, \&
  {Rousset}}]{mazoyer14a}
{Mazoyer}, J., {Baudoz}, P., {Galicher}, R., \& {Rousset}, G. 2014, Astronomy
  and Astrophysics, 564, L1

\bibitem[{{Murakami} {et~al.}(2008){Murakami}, {Uemura}, {Baba}, {Nishikawa},
  {Tamura}, {Hashimoto}, \& {Abe}}]{murakami08}
{Murakami}, N., {Uemura}, R., {Baba}, N., {et~al.} 2008, Publication of the
  Astronomical Society of Pacific, 120, 1112

\bibitem[{{Patru} {et~al.}(2018){Patru}, {Baudoz}, {Galicher}, {Boussaha},
  {Firminy}, {Cao}, {Wang}, {Xing}, {Bonafous}, \& {Potier}}]{patru18}
{Patru}, F., {Baudoz}, P., {Galicher}, R., {et~al.} 2018, in Society of
  Photo-Optical Instrumentation Engineers (SPIE) Conference Series, Vol. 10703,
  Adaptive Optics Systems VI, 107032L

\bibitem[{{Potier} {et~al.}(2018){Potier}, {Baudoz}, {Galicher}, {Patru}, \&
  {Thijs}}]{potier18}
{Potier}, A., {Baudoz}, P., {Galicher}, R., {Patru}, F., \& {Thijs}, S. 2018,
  in Society of Photo-Optical Instrumentation Engineers (SPIE) Conference
  Series, Vol. 10698, Proceedings of the SPIE, 106986G

\bibitem[{Racine {et~al.}(1999)Racine, Walker, Nadeau, Doyon, \&
  Marois}]{racine99}
Racine, R., Walker, G.~A.~H., Nadeau, D., Doyon, R., \& Marois, C. 1999,
  Publications of the Astronomical Society of the Pacific, 111, 587

\bibitem[{{Raymond} {et~al.}(2014){Raymond}, {Kokubo}, {Morbidelli},
  {Morishima}, \& {Walsh}}]{raymond14a}
{Raymond}, S.~N., {Kokubo}, E., {Morbidelli}, A., {Morishima}, R., \& {Walsh},
  K.~J. 2014, 595

\bibitem[{{Rouan}(2016)}]{rouan16}
{Rouan}, D. 2016, in EAS Publications Series, Vol. 78-79, 73--98

\bibitem[{Rouan {et~al.}(2000)Rouan, Riaud, Boccaletti, Cl{\'e}net, \&
  Labeyrie}]{Rouan00}
Rouan, D., Riaud, P., Boccaletti, A., Cl{\'e}net, Y., \& Labeyrie, A. 2000,
  Publications of the Astronomical Society of the Pacific, 112, 1479

\bibitem[{Santos(2008)}]{santos08}
Santos, N.~C. 2008, New Astronomy Review, 52, 154

\bibitem[{{Santos} {et~al.}(2017){Santos}, {Adibekyan}, {Figueira},
  {Andreasen}, {Barros}, {Delgado-Mena}, {Demangeon}, {Faria}, {Oshagh},
  {Sousa}, {Viana}, \& {Ferreira}}]{santos17}
{Santos}, N.~C., {Adibekyan}, V., {Figueira}, P., {et~al.} 2017, \aap, 603, A30

\bibitem[{{Santos} \& {Faria}(2018)}]{santos18}
{Santos}, N.~C. \& {Faria}, J.~P. 2018, in Asteroseismology and Exoplanets:
  Listening to the Stars and Searching for New Worlds, Vol.~49, 165

\bibitem[{{Serabyn} {et~al.}(2019){Serabyn}, {Prada}, {Chen}, \&
  {Mawet}}]{serabyn19}
{Serabyn}, E., {Prada}, C.~M., {Chen}, P., \& {Mawet}, D. 2019, Journal of the
  Optical Society of America B Optical Physics, 36, D13

\bibitem[{{Singh} {et~al.}(2019){Singh}, {Galicher}, {Baudoz}, {Dupuis},
  {Ortiz}, {Potier}, {Thijs}, \& {Huby}}]{singh19}
{Singh}, G., {Galicher}, R., {Baudoz}, P., {et~al.} 2019, Astronomy and
  Astrophysics paper, 631, A106

\bibitem[{{Singh} {et~al.}(2014){Singh}, {Guyon}, {Baudoz}, {Jovanovic},
  {Martinache}, {Kudo}, {Serabyn}, \& {Kuhn}}]{singh14}
{Singh}, G., {Guyon}, O., {Baudoz}, P., {et~al.} 2014, Submitted

\bibitem[{{Snellen} {et~al.}(2014){Snellen}, {Brandl}, {de Kok}, {Brogi},
  {Birkby}, \& {Schwarz}}]{snellen14}
{Snellen}, I. A.~G., {Brandl}, B.~R., {de Kok}, R.~J., {et~al.} 2014, Nature,
  509, 63

\bibitem[{{Snellen} {et~al.}(2010){Snellen}, {de Kok}, {de Mooij}, \&
  {Albrecht}}]{snellen10}
{Snellen}, I. A.~G., {de Kok}, R.~J., {de Mooij}, E. J.~W., \& {Albrecht}, S.
  2010, Nature, 465, 1049

\bibitem[{{von Essen} {et~al.}(2019){von Essen}, {Mallonn}, {Welbanks},
  {Madhusudhan}, {Pinhas}, {Bouy}, \& {Weis Hansen}}]{essen19}
{von Essen}, C., {Mallonn}, M., {Welbanks}, L., {et~al.} 2019, Astronomy and
  Astrophysics, 622, A71

\end{thebibliography}

\end{document}